\documentclass{pasj00}
\draft

\begin{document}
\SetRunningHead{T. Anada et al.}{X-Ray Studies of HESS J1809--193 with Suzaku}
%\Received{2000/0/0}%{yyyy/mm/dd}
%\Accepted{2000/0/0}%{yyyy/mm/dd}

\title{X-Ray Studies of HESS J1809--193 with Suzaku}

%%% begin:list of authors
% Do NOT capitalize all letters in "textsc".
\author{Takayasu \textsc{Anada}\altaffilmark{1},
Aya {\sc Bamba}\altaffilmark{2,1},\thanks{Corresponding author: abamba@cp.dias.ie},
Ken \textsc{Ebisawa}\altaffilmark{1},
and
Tadayasu {\sc Dotani}\altaffilmark{1}
}
\altaffiltext{1}{Institute of Space and Astronautical Science, JAXA, 3-1-1
Yoshinodai, Sagamihara, Kanagawa 229-8510}
\altaffiltext{2}{School of Cosmic Physics,
Dublin Institute for Advanced Studies
31 Fitzwilliam Place, Dublin 2,
Ireland
}
%\email{anada@astro.isas.jaxa.jp, ebisawa@astro.isas.jaxa.jp,
%dotani@astro.isas.jaxa.jp, bamba@astro.isas.jaxa.jp}
%%% end:list of authors

\maketitle

\begin{abstract}
Suzaku observed the region including HESS~J1809$-$193,
one of the TeV unidentified (unID) sources, and confirmed
existence of the extended hard X-ray emission previously reported by ASCA, 
as well as hard X-ray emission from the pulsar PSR~J1809$-$1917 
in the region.
One-dimensional profile of the diffuse emission is represented with
a Gaussian model with the best-fit $\sigma$ of $7\pm1$ arcmin.
The diffuse emission extends for at least 21~pc
(at the 3$\sigma$ level, assuming the distance of 3.5~kpc),
and has a hard spectrum with the photon index of $\Gamma \sim$1.7.
%compared to
%those of non-thermal supernova remnants ($\Gamma \sim$2 -- 3).
The  hard spectrum  suggests the pulsar wind nebula (PWN) origin, 
which is also strengthened by the hard X-ray emission from 
PSR~J1809$-$1917 itself.
Thanks to the low background of Suzaku XIS, we were able to investigate
 spatial variation of the energy spectrum, but 
no systematic spectral change in the extended emission is found.
These results imply that
the X-ray emitting pulasr wind electrons
can travel up to 21~pc from the pulsar
without noticable energy loss via synchrotron emission.
\end{abstract}

\section{Introduction}

Galactic plane survey with the H.E.S.S. Cherenkov telescope system
revealed % presence of 
dozens of the new very-high-energy (VHE) $\gamma$-ray
sources~\citep{2005Sci...307.1938A,2006ApJ...636..777A}. Many of them
have no counterparts in other wave-lengths,  thus called
``unidentified (unID) TeV sources''.
Today, about 40 such unidentified TeV sources are known on the
Galactic plane~\citep{2007arXiv0712.3352H}. Most of them are located
within a height of $\pm$ 1 degree from the Galactic plane,
and some are intrinsically extended.
Despite a large number of intensive studies in the last several years, 
their origin is unclear \citep{2007arXiv0712.3352H}.

X-ray follow-up observations of the unID TeV sources are now on-going.
% with enthusiasm.
Although supernova remnants (SNRs) or hypernova remnants were suggested to be
major counterpart candidates of these TeV unID sources
\citep{yamazaki2006,ioka2009},
only a few sources have been actually identified as SNRs
\citep{2009PASJ...61S.197N,2008A&A...477..353A}.
On the other hand, rather surprisingly, 
several unID TeV sources have been identified as pulsar wind nebulae (PWNe)
\citep[ for example]{2009PASJ...61S.183A,2009PASJ...61S..189T}.
%implying that
% the main component of TeV unID sources is, suprisingly, PWNe.
They seem to be rather old, previously unknown PWNe, compared to the PWNe 
already identified and  well studied in X-rays.

The first HESS observations of the region around PSR J1809--1917 were
made from May through 
June 2004 as part of the systematic survey of the inner Galaxy
\citep{2005Sci...307.1938A,2006ApJ...636..777A}.
Because
 marginal VHE $\gamma$-ray signals were detected, HESS J1809--193 was
observed again in 2004 and 2005, and significant $\gamma$-ray emission
was confirmed~\citep{2007A&A...472..489A}. Recent study of this source by HESS
was reported by \citet{2008AIPC.1085..285R}. Fitting the excess map with a 2-D
symmetric Gaussian, the best fit position and intrinsic source extension
(in rms) were determined as (RA, Dec) =
($18^{\mathrm{h}}09^{\mathrm{m}}52^{\mathrm{s}}$, $-19^{\circ}23'42''$)
and $0^\circ.25\pm0^\circ.02$, respectively.

PSR J1809--1917 is a radio pulsar discovered by the Parkes Multibeam
Pulsar Survey~\citep{2002MNRAS.335..275M}. The pulsar is located at the
position of (RA, Dec) = (\timeform{18h09m43s.1}, \timeform{-19D17'38"}) 
with a pulse period of $P=82.7$~ms and the period
derivative of $\dot{P} = 2.55\times10^{-14}$~s~s$^{-1}$. The distance to
the source was estimated to be $d=3.5$~kpc from the pulsar's dispersion
measure using the NE2001 Galactic electron-density
model~\citep{2002astro.ph..7156C}. The characteristic age and the
spin-down luminosity are $\tau {c}=51$~kyr and $\dot{E} =
1.8\times10^{36}$~ergs~s$^{-1}$, respectively.

The $\gamma$-ray spectral analysis performed by \citet{2007arXiv0709.2432K}
indicated that the spectral slope is different between the regions near
the pulsar and far from the pulsar.
This is the second case that  spatial variation of the spectral slope 
is  revealed in the VHE $\gamma$-ray emission.
The first case is HESS J1825--137~\citep{2006A&A...460..365A}, 
which is largely extended in both VHE $\gamma$-ray and X-ray 
bands~\citep{2009PASJ...61S..189T}.

ASCA observation revealed diffuse, non-thermal emission in the vicinity of PSR
J1809--1917~\citep{2003ApJ...589..253B}. \citet{2007ApJ...670..655K}
detected a bright point X-ray source which was positionally consistent
with the pulsar PSR J1809$-$1917, and resolved the surrounding compact PWN
utilizing the very high angular resolution of Chandra. The 
PWN has a ``head-tail'' profile, consisting of the southern-head, which is 
coincident to the  pulsar,  and the northern-tail.
\citet{2007ApJ...670..655K} claimed that this cometary morphology is 
attributed to a bow shock
created by the pulsar moving supersonically to the southern direction.
However, it is still unknown whether the faint diffuse emission 
discovered by ASCA is related to the PWN\@ or not.
In this paper, we make detailed analysis of the diffuse emission
for the first time using Suzaku.  Suzaku,  characterized by the low detector
background compared to Chandra and XMM-Newton, is  very  suitable for the analysis of faint and diffuse X-ray emission such as HESS 1809--193.

\section{Observations}
\label{sec:observation}

\begin{figure}[htbp]
\begin{center}
\FigureFile(80mm,80mm)
{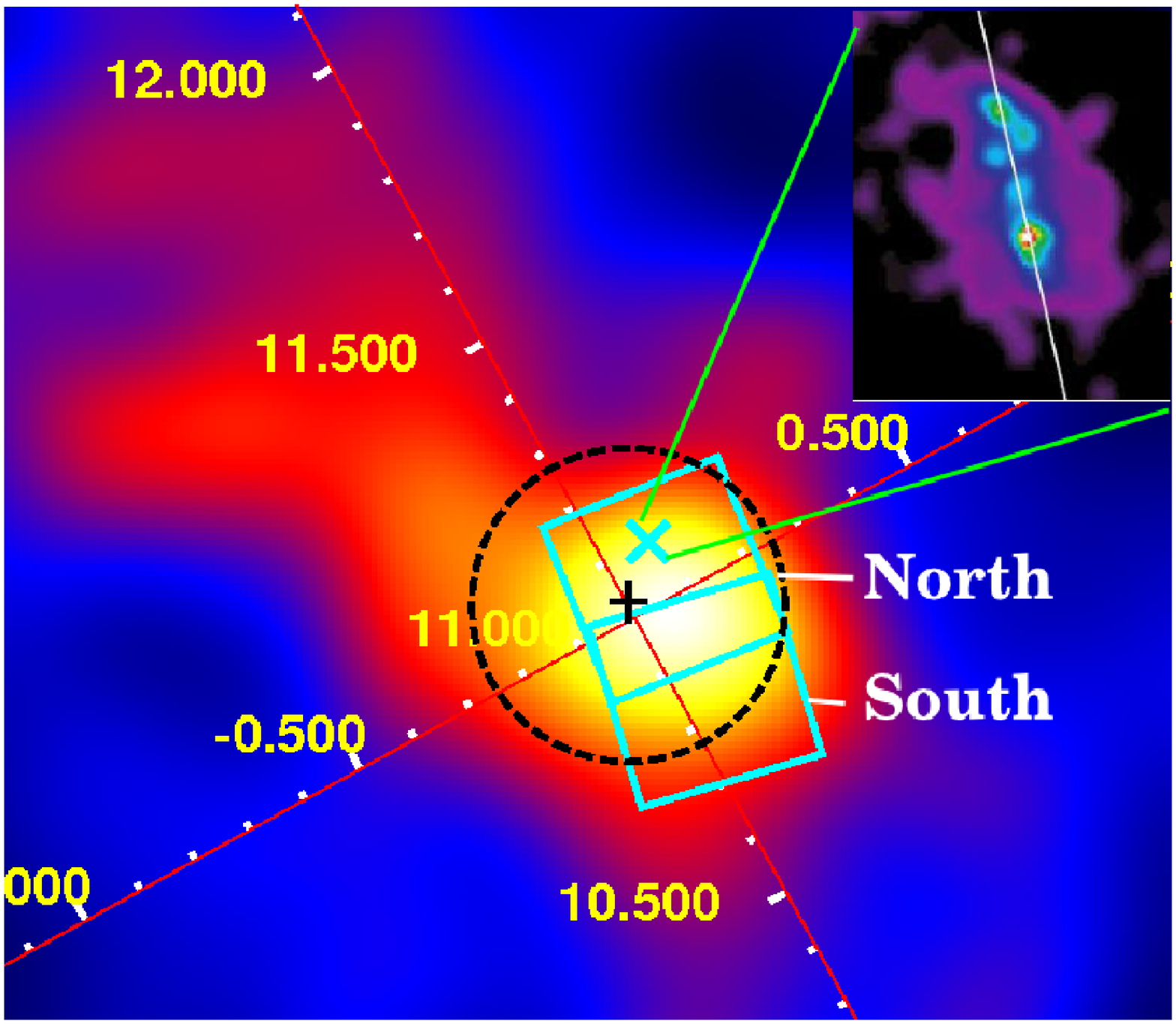}
\end{center}
\caption{
A smoothed excess map measured by HESS around HESS J1809--193 
\citep{2007A&A...472..489A}.
%Fits image
%was provided by Gerd P$\ddot{\mathrm{u}}$hlhofer.
Position of the
pulsar PSR J1809--1917 is marked with a cyan cross,
whose Chandra image \citep{2007ApJ...670..655K} is shown in the inset
at the top right of the figure.
The best fit position and extent of the $\gamma$-ray source is marked with
a black cross and a dashed circle, respectively.
Two cyan squares indicate the two XIS field of views (FOVs) during the Suzaku
observations, which are referred to as 
the North and the South pointing, respectively.
}
\label{fig:1809:hessj1809}
\end{figure}

We observed HESS J1809--193 with Suzaku
\citep{mitsuda2007}
in April, 2008.
Suzaku is equipped with two types of instruments: the X-ray Imaging Spectrometers
(XIS:~\cite{2007PASJ...59S..23K}) at the foci of four X-Ray Telescopes
(XRT:~\cite{2007PASJ...59S...9S}) and the Hard X-ray Detector
(HXD:~\cite{2007PASJ...59S..35T}, \cite{2007PASJ...59S..53K}).
The observation was carried out with two pointings at north and
south of the source region (figure \ref{fig:1809:hessj1809}) in order to cover the pulsar and extended VHE $\gamma$-ray
emission along the direction of the elongated shape of the PWN \citep{2007ApJ...670..655K}.
Three XISs (XIS 0, 1, 3) out of four were operated in the normal clocking mode with
the Spaced-row Charge Injection (SCI)~\citep{2008PASJ...60S...1N}.
We analyzed the data processed by the version 2.2 pipeline.
We are interested in the spatial variations of a scale of arcminutes.
Hence, we concentrated on the analysis of the XIS data in this paper, since
 the HXD does not have a
spatial resolution within  a field of view of $\sim$30~arcmin.\ (FWHM)\@.
We applied the standard screening criteria to the XIS 
data\footnote{http://www.astro.isas.jaxa.jp/suzaku/process/v2changes/criteria\ xis.html}
to obtain the cleaned event lists.
After the data screening, the net exposures was 51.5~ks and 44.2~ks for
the north and south pointing of XIS, respectively.
The Suzaku observation log and exposure are summarized in table~\ref{tbl:1809:obslog}.
We used HEADAS version 6.5 software package for the data analysis.

\begin{table*}
\begin{center}
\caption{Journal of the Suzaku observations of HESS
J1809--193}
\label{tbl:1809:obslog}
\begin{tabular}{ccc}
\hline
& North & South \\
\hline
Sequence ID & 503078010 & 503079010 \\
Start time (UT)\footnotemark[1] & 2008/03/31 14:06 & 2008/04/01 16:34 \\
End time (UT)\footnotemark[1] & 2008/04/01 16:30 & 2008/04/02 14:47 \\
Aim point R.A. (J2000.0) &
$18^{\mathrm{h}}09^{\mathrm{m}}37^{\mathrm{s}}.4$ &
$18^{\mathrm{h}}09^{\mathrm{m}}21^{\mathrm{s}}.0$ \\
Aim point Decl. (J2000.0) &
$-19^{\circ}21'24''$ &
$-19^{\circ}32'02''$ \\
Net exposure (ks) & 51.5 & 44.2 \\
\hline
\multicolumn{3}{@{}l@{}}{\hbox to 0pt{\parbox{180mm}{\footnotesize
\footnotemark[1] Time form of yyyy/mm/dd hh:mm
}\hss}}
\end{tabular}
\end{center}
\end{table*}

\section{Results}
\label{sec:results}

\subsection{X-ray Image}
\label{sec:image}

\begin{figure}[tbp]
\begin{center}
\FigureFile(60mm,60mm)
{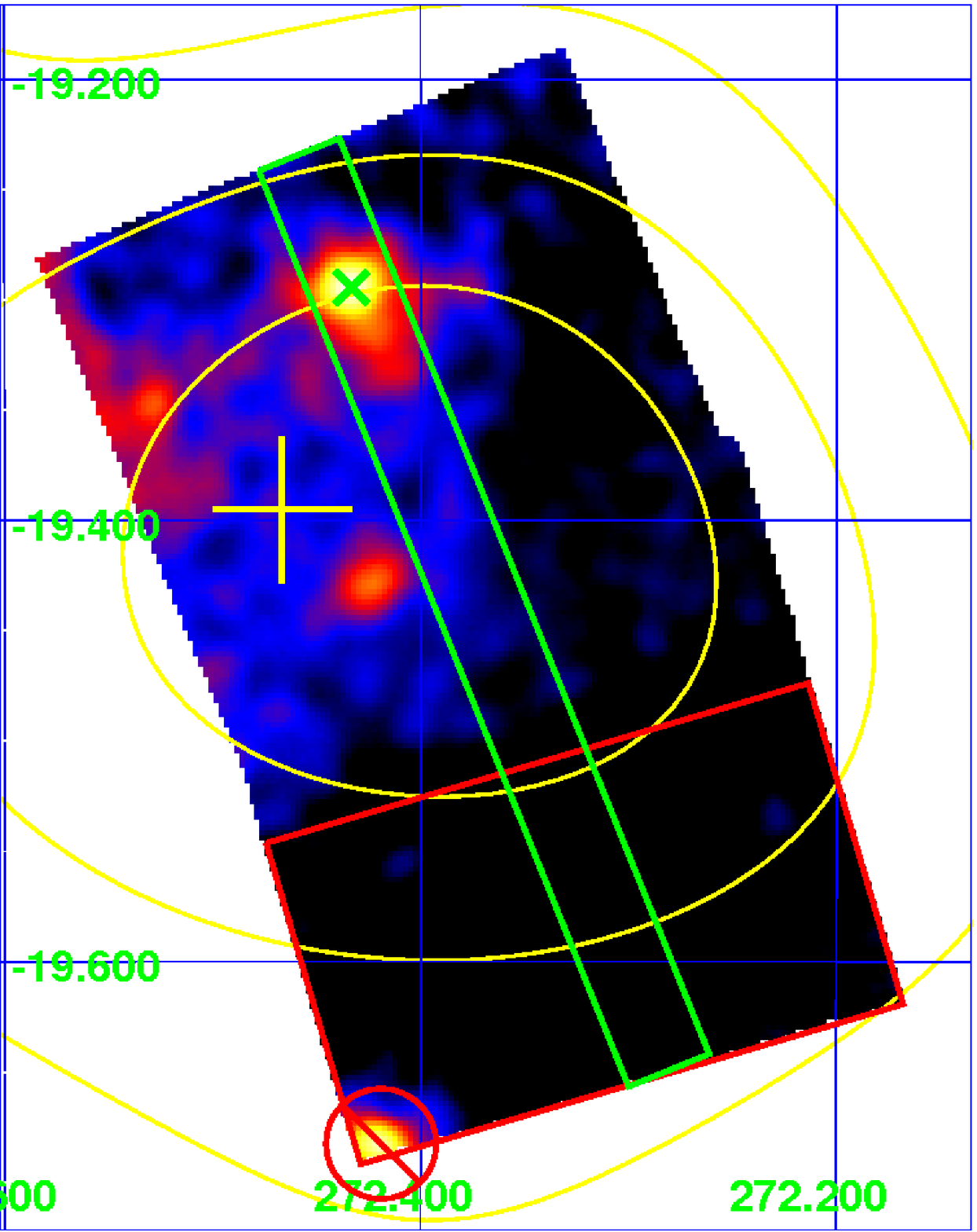}
\FigureFile(60mm,60mm)
{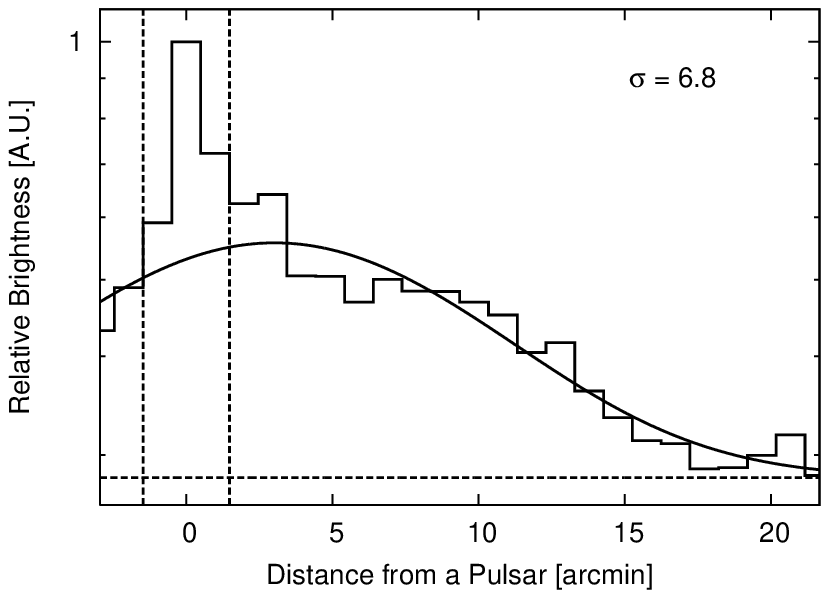}
\end{center}
\caption{
Left: Suzaku XIS image of the region including HESS J1809--193.
The pseudo-color represents vignetting-corrected, log-scaled intensity
levels in the 2.0--10.0~keV band. 
The yellow cross indicates the centroid of HESS J1809--193, and the yellow lines represent 80\%,
60\%, and 40\% contour of the peak of the VHE $\gamma$-ray~\citep{2008AIPC.1085..285R}.
Right: normalized 1-dimensional profile of the surface brightness in
the 2.0--10.0~keV band along the direction from north to south in the
green rectangle indicated in figure~\ref{fig:1809:img reg-hess} with a
spatial bin size of $1'.0$. Solid curve and horizontal dashed line show
the best-fit Gaussian profile and a constant, respectively. The bright peak
corresponding to the pulsar was ignored in this fit ($2'.9$ wide, inner part
between the two vertical dashed lines).
}
\label{fig:1809:img reg-hess}
\end{figure}

In order to make images of this field,
we corrected the vignetting effect
by dividing the image by the flat sky image
simulated using {\tt xissim} \citep{2007PASJ...59S.113I}
after subtracting non X-ray background \citep{2008PASJ...60S..11T}.
In this simulation,
we assumed the input energy spectrum as that extracted from
the red rectangular region (see \S \ref{sec:1809:ana:spec}). 
Hereafter, all images are vignetting corrected.
Figure~\ref{fig:1809:img reg-hess} shows Suzaku XIS 2--10~keV image
of the HESS~J1809--193 region.% after the all corrections.

We confirmed largely extended emission
reported by ASCA \citep{2003ApJ...589..253B}. In addition, we found 
the extended emission has several peaks.
The pulsar is the brightest in the FOV, 
and the western edge of the FOV is also significantly bright, 
which was already hinted by ASCA \citep{2003ApJ...589..253B}.

We have determined extension of the diffuse emission
as follows:
We created a 1-dimensional profile of the surface brightness from
the rectangle region shown in
figure~\ref{fig:1809:img reg-hess} (left)
which runs from north to south.
% Because we are interested in the faint and extended emission, 
We selected the direction which enabled us to create the longest profile.
We did not  take a symmetric region around the pulsar, 
because the faint and diffuse emission is  significantly asymmetric.
Although there is a point source in the south of the pulsar at
(272.42, $-$19.43),
its flux in the integrated region is
only 1 \% of the central pulsar, thus negligible.
The 1-dimensional profile thus created is shown in
figure~\ref{fig:1809:img reg-hess} (right). Note that the surface
brightness is normalized to the peak brightness. We fitted the profile
with a Gaussian function plus a constant to evaluate its extension.
In the fitting, we ignored the brightest part 
around the pulsar with a width of $2'.9$ (corresponding to the
point-spread function of the XRT)\@.
Consequently, we found the diffuse emission extends up to $\sim20'$ away from the pulsar.
The Gaussian center was found to be offset by $\sim\!3'$ from the pulsar
and the rms width to be $\sigma=7'\pm1'$.

\subsection{Energy Spectra}
\label{sec:1809:ana:spec}

We studied spatial variation of the X-ray energy spectra of
the diffuse emission.
We generated the  detector response file and an auxiliary response file
using {\tt xisrmfgen},  and {\tt xissimarfgen} \citep{2007PASJ...59S.113I}
and performed model fitting using XSPEC version 12.4.0.

\begin{figure}[tbp]
\begin{center}
\FigureFile(80mm,80mm)
{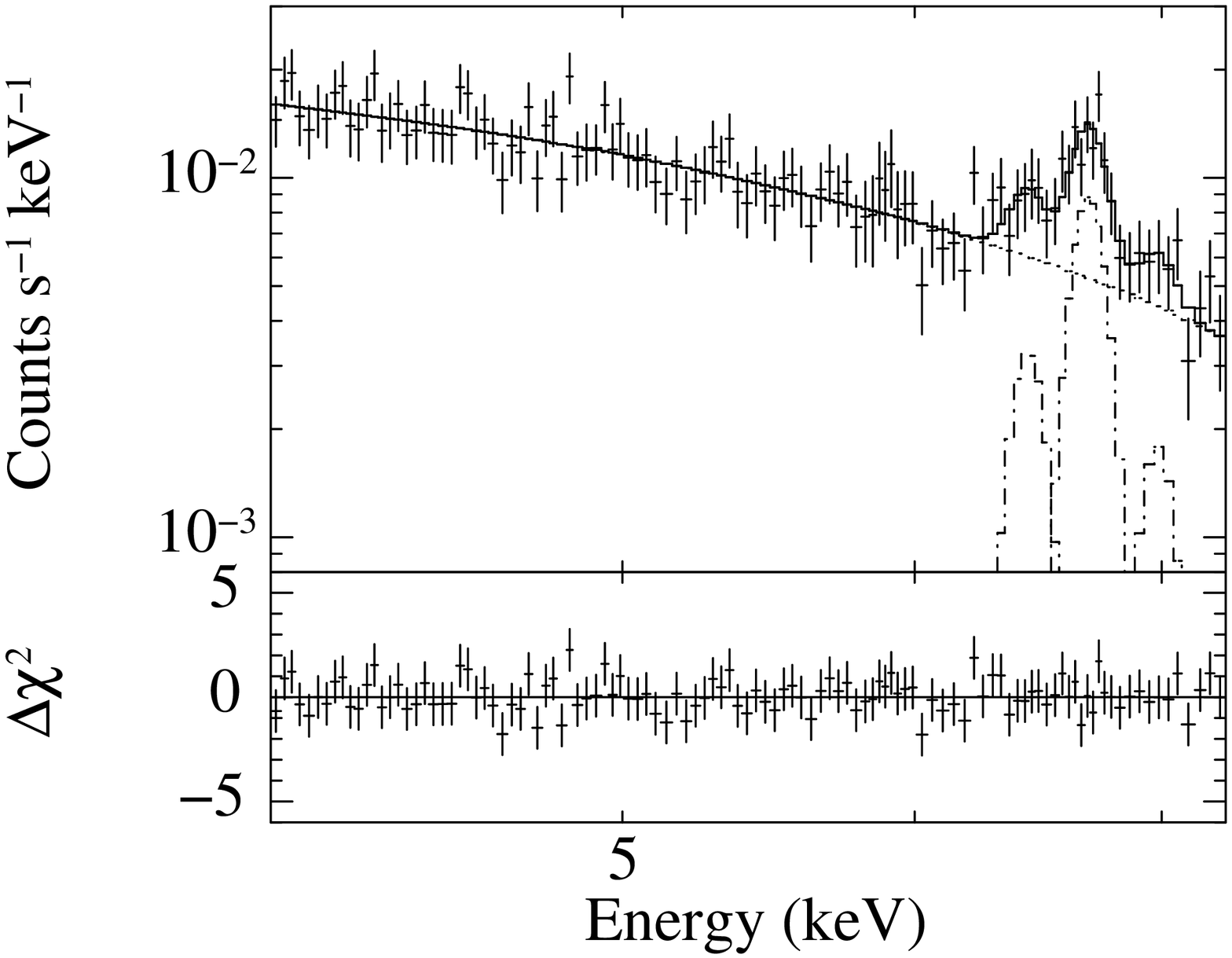}
\hspace{0.5cm}
\FigureFile(80mm,80mm)
{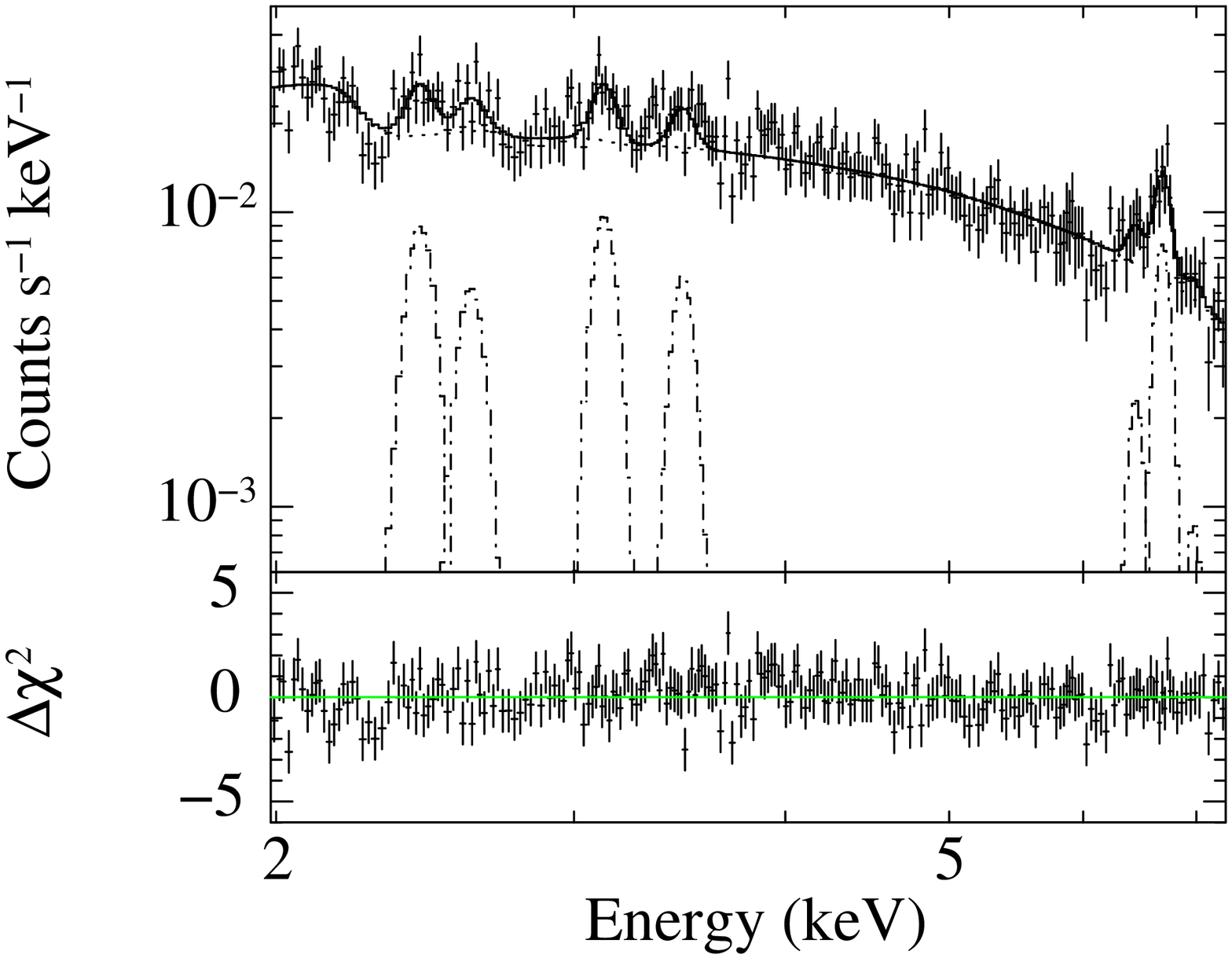}
\end{center}
\caption{
Left: NXB-subtracted XIS spectrum in the 4.0--7.2~keV band extracted
from the region enclosed with a red rectangle in figure \ref{fig:1809:img reg-hess} (left). The best-fit model, an
absorbed power-law with three iron lines, is also shown in a solid line
(each component in a dot-dashed line). The bottom panel shows residuals
to the best-fit model. Right: the same as the left one, but in the
2.0--7.2~keV band. Narrow Gaussians were added to the model below
4~keV.
}
\label{fig:1809:bgd spec}
\end{figure}

First we examined energy spectrum from an outskirt of the
HESS source region (the red rectangular region in
figure~\ref{fig:1809:img reg-hess}) in order to estimate contribution of the
Galactic Ridge X-ray Emission (GRXE) on this particular area.
The bright source at the south-east corner (circled in red) was excluded in this
analysis. We subtracted the non X-ray background
(NXB) estimated using {\tt xisnxbgen} (detail of the method described
in \cite{2008PASJ...60S..11T}), and
averaged all of the available XISs (XIS 0,1,3) data. 
Because the NXB
becomes high above 7.2~keV (in particular for XIS1 which has the back-illuminated chip),
we used the data only below 7.2~keV. We fitted the spectra in 
4.0--7.2 keV and 2.0--4.0~keV separately following the reproduction procedure of 
GRXE in \citet{2008PASJ...60S.223E}.
% since we are mostly interested in the background GRXE.
The model adopted is a power-law plus three narrow
Gaussian (intrinsic width fixed to zero) % , and the center energies
% optimized around 6.4~keV, 6.7~keV and 7.0~keV)\@ 
 in 4.0 -- 7.2 keV, or
four gaussian lines in 2.0 -- 4.0 keV. The spectra and the
best-fit models are shown in figure~\ref{fig:1809:bgd spec}. 
The equivalent widths of the iron lines are summarized in
table~\ref{tbl:1809:eqwidth}. 
These equivalent widths are comparable to
those of GRXE described in \citet{2008PASJ...60S.223E}. Thus we conclude
that the X-ray emission in this region can be regarded as pure GRXE and
no significant contribution from the pulsar is present.

\begin{table*}[tbp]
\begin{center}
\caption{Center energies and equivalent widths of the iron lines in the background region in comparison of those of the GRXE. }
\label{tbl:1809:eqwidth}
\begin{tabular}{lccc}
\hline\hline
\multicolumn{4}{c}{Background region} \\
\hline
Center energy (keV) & $6.44\pm0.07$ & $6.69\pm0.02$ & $6.98\pm0.08$ \\
Equivalent width (eV) & 70 (0--140) & 240 (160--550) & 50 (0--170) \\
\hline\hline
\multicolumn{4}{c}{GRXE$^{*}$} \\
\hline
Center Energy (keV) & $6.41\pm0.02$ & $6.670\pm0.006$ & $7.00\pm0.03$ \\
Equivalent width (eV) $\quad$ & 80 (60--100) & 350 (310--390) & 70 (40--100) \\
\hline
\multicolumn{4}{@{}l@{}}{\hbox to 0pt{\parbox{160mm}{\footnotesize
Note. -- Errors represent single-parameter 90\% confidence limit.
\par\noindent
$^{*}$ Center energies and equivalent widths determined by the GRXE observation with
Suzaku by \citet{2008PASJ...60S.223E}.
\par\noindent
}\hss}}
\end{tabular}
\end{center}
\end{table*}

\begin{figure}[tbp]
\begin{center}
\FigureFile(80mm,80mm)
{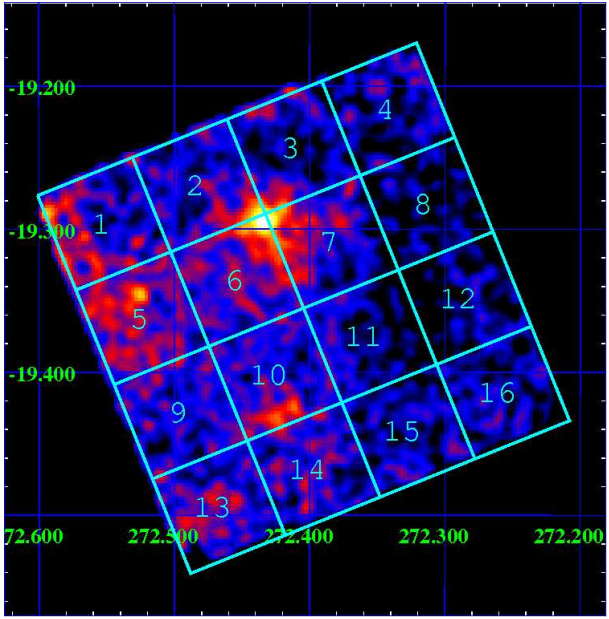}
\hspace{0.5cm}
\FigureFile(80mm,80mm)
{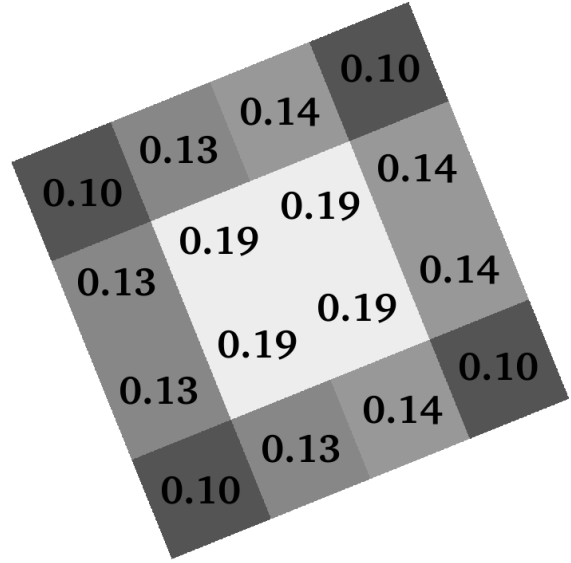}
\end{center}
\caption{
Left: Suzaku XIS image of the north pointing in the 2.0-10.0 keV band
and designation of the 16 sub-regions for spectral variation study.
Right: Correction factors for each region, which take account of the 
vignetting effects,  differences of the exposure and
the extraction area between the source region and the GRXE region 
(the red rectangle in figure~\ref{fig:1809:img reg-hess}).
}
\label{fig:1809:sum013 reg 0548-2740 expcor}
\end{figure}

In the next step,
we divided the sky region of the north pointing into
a ``check pattern'' as shown in
figure~\ref{fig:1809:sum013 reg 0548-2740 expcor} (left)
in order to find out
the possible spatial variations of the spectral slope. 
Here we refer to
these regions as the number indicated in
figure~\ref{fig:1809:sum013 reg 0548-2740 expcor} (left) with the prefix
``Grid'' (Grid1, Grid2, ...). We used a two component model for the fit:
an absorbed power-law plus the GRXE\@.
%Because of the poor statistics of data in each grid, 
We used the model
GRXE spectrum (as explained below)  % instead of the observed GRXE spectrum 
% from the background region in figure~\ref{fig:1809:img reg-hess}
to subtract the diffuse background.

\begin{figure}[tbp]
\begin{center}
\FigureFile(80mm,80mm)
{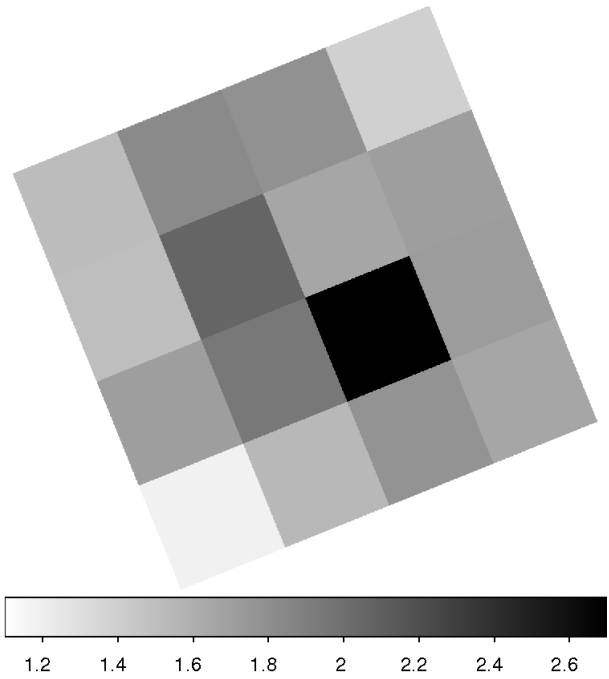}
\hspace{0.5cm}
\FigureFile(80mm,80mm)
{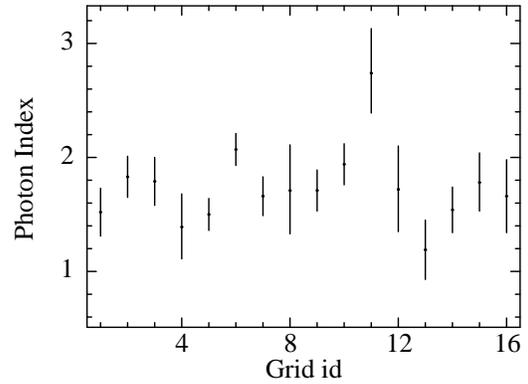}
\end{center}
\caption
{Spatial distribution of the spectral indices around PSR
J1809--1917. Left: best-fit photon index map. Gray scale indicates the
photon indices; light gray means the spectrum is hard while dark gray
means soft. Right: photon indices of each grid with 90\% error bars.
}
\label{fig:1809:phoindex grid}
\end{figure}

\begin{figure}[tbp]
\begin{center}
% \FigureFile(160mm,160mm)
% {fig/multi sum013 gridall min50 fit.pha.eps}
\FigureFile(40mm,40mm){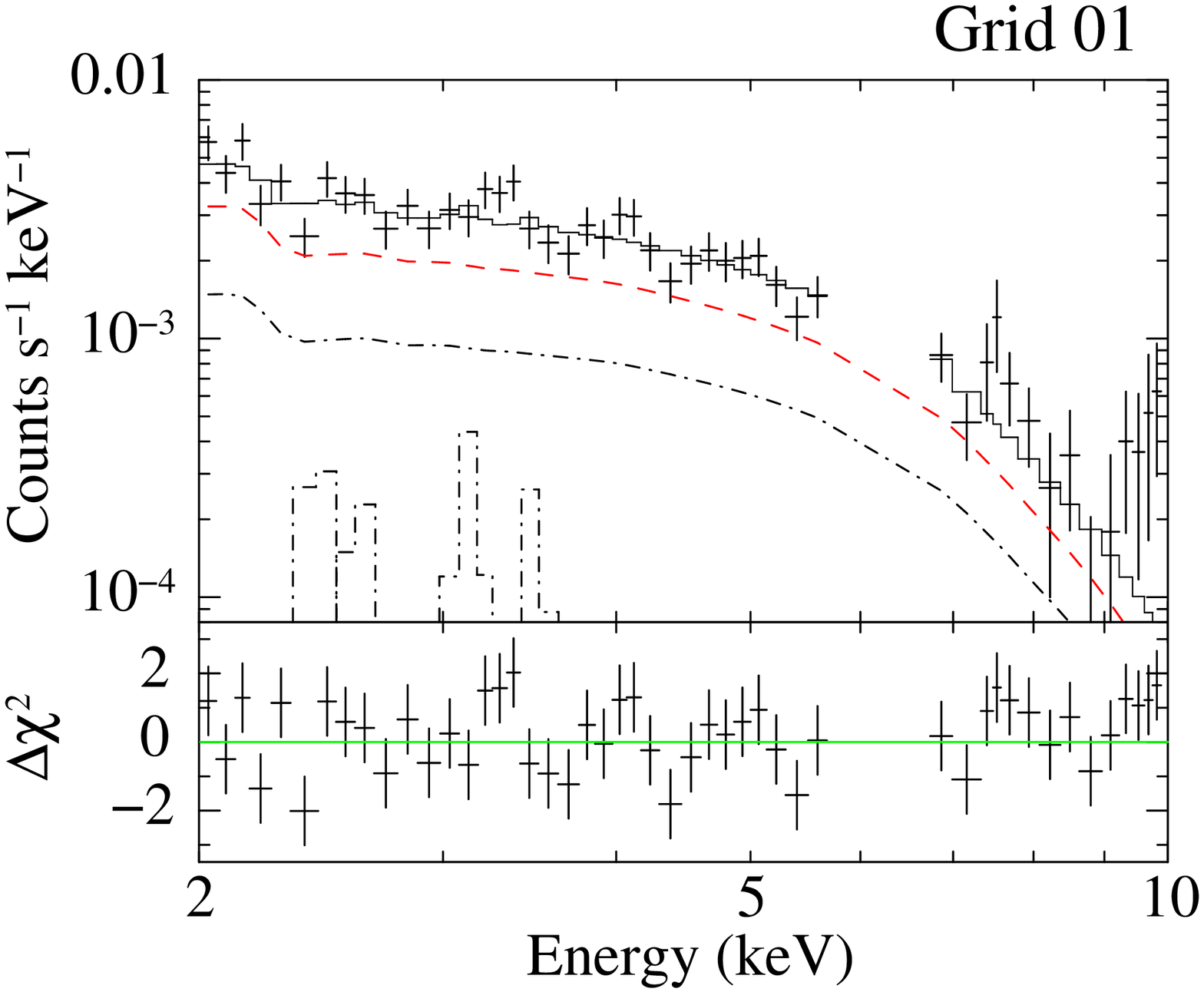}
\FigureFile(40mm,40mm){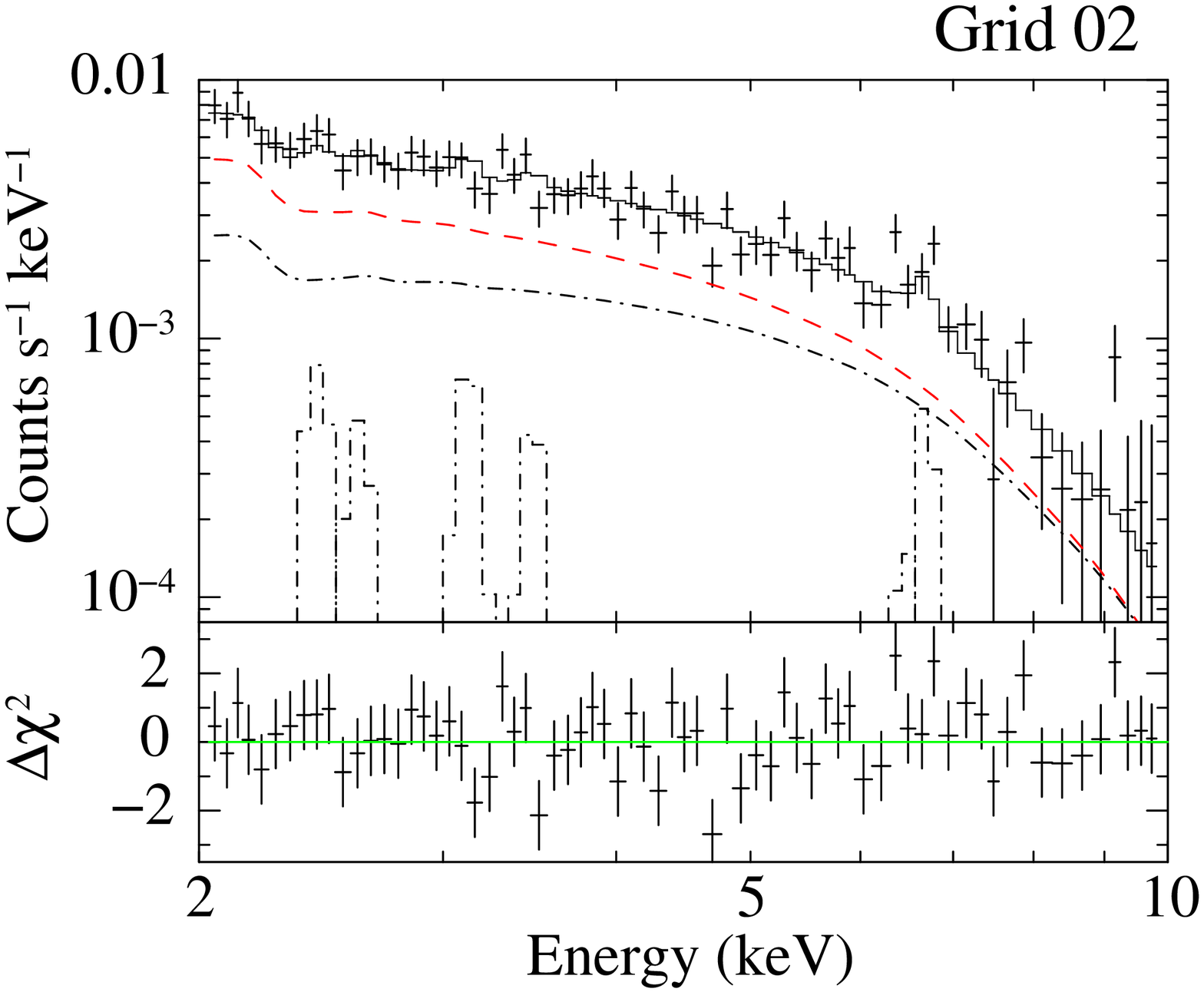}
\FigureFile(40mm,40mm){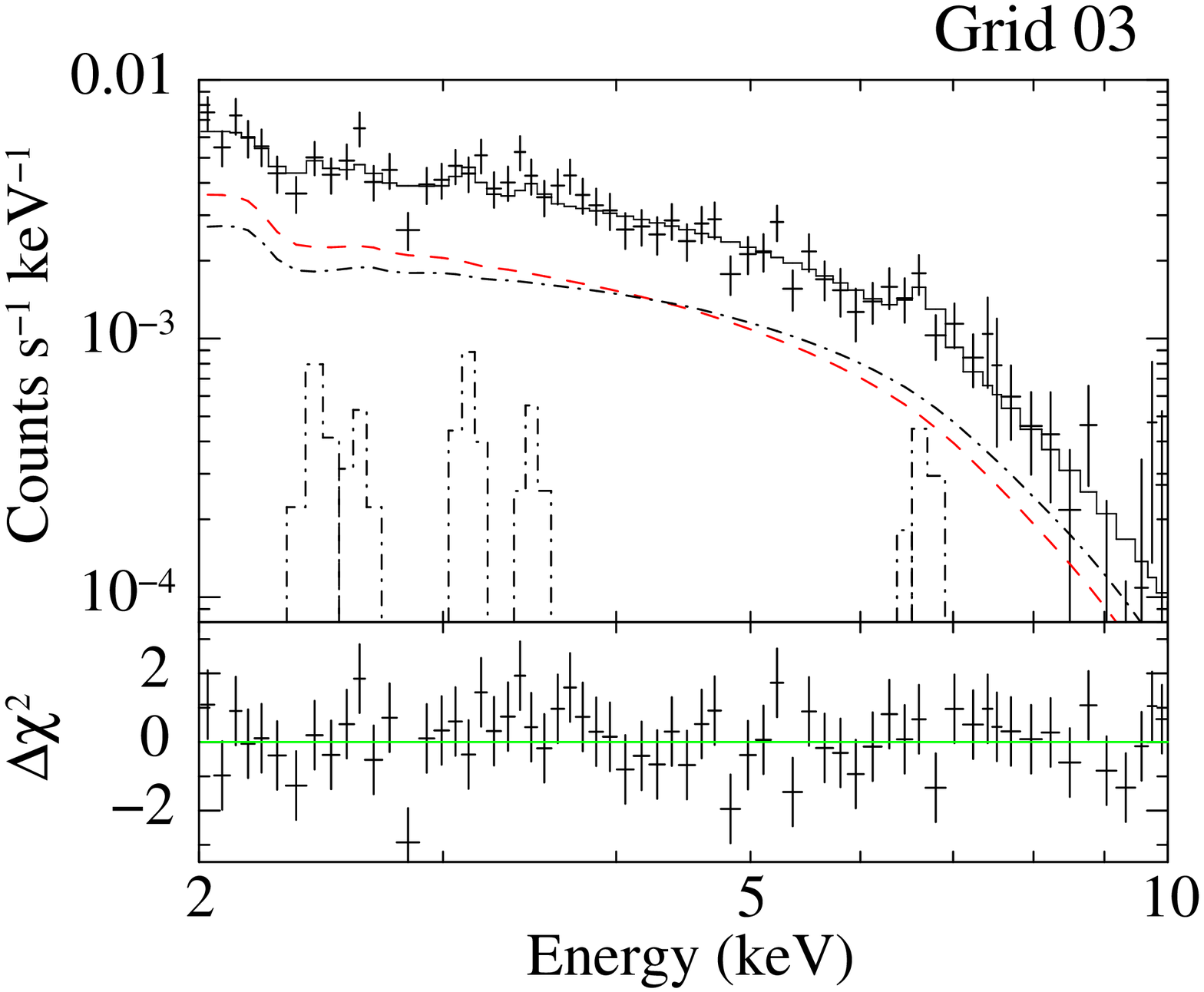}
\FigureFile(40mm,40mm){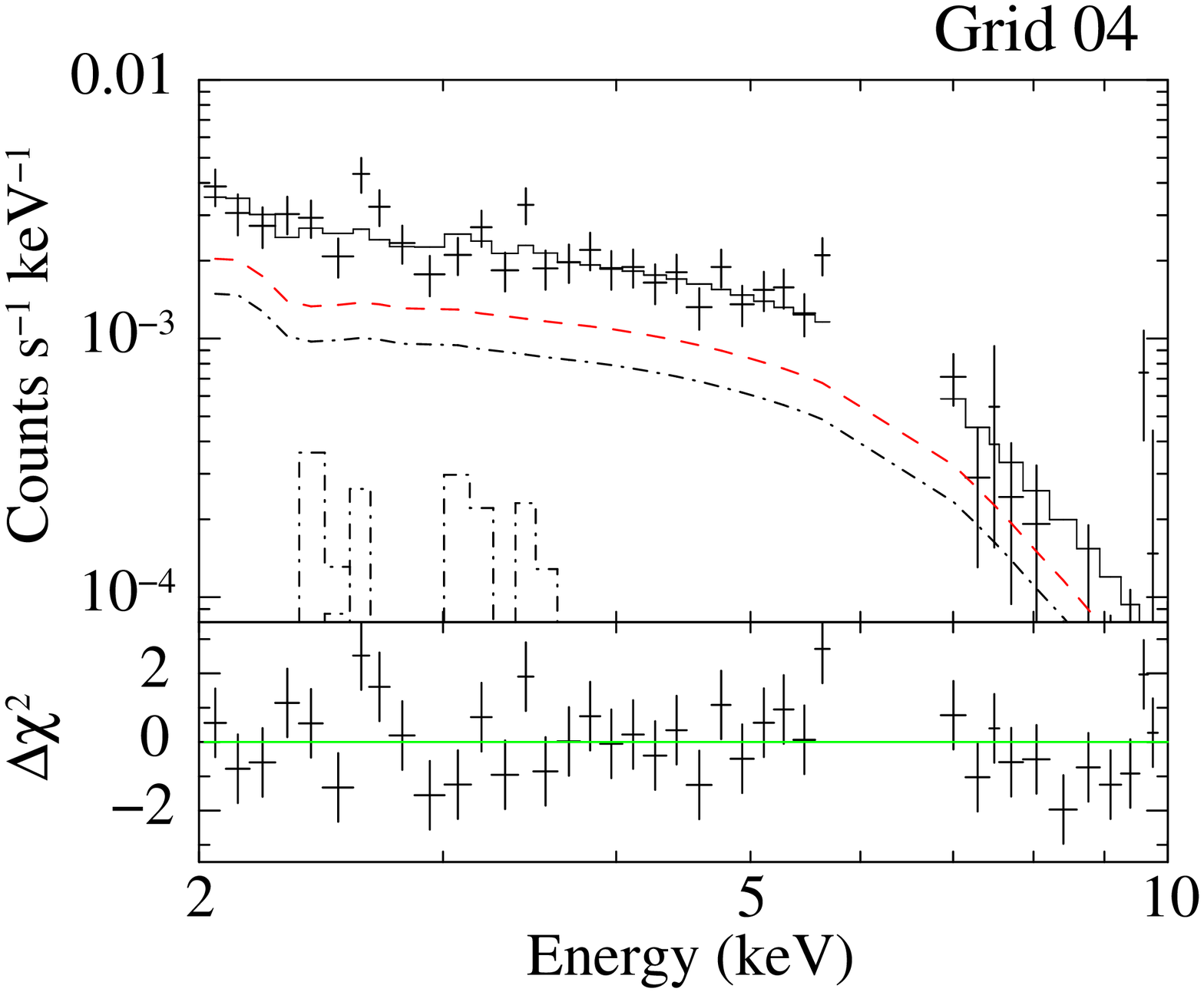}
\FigureFile(40mm,40mm){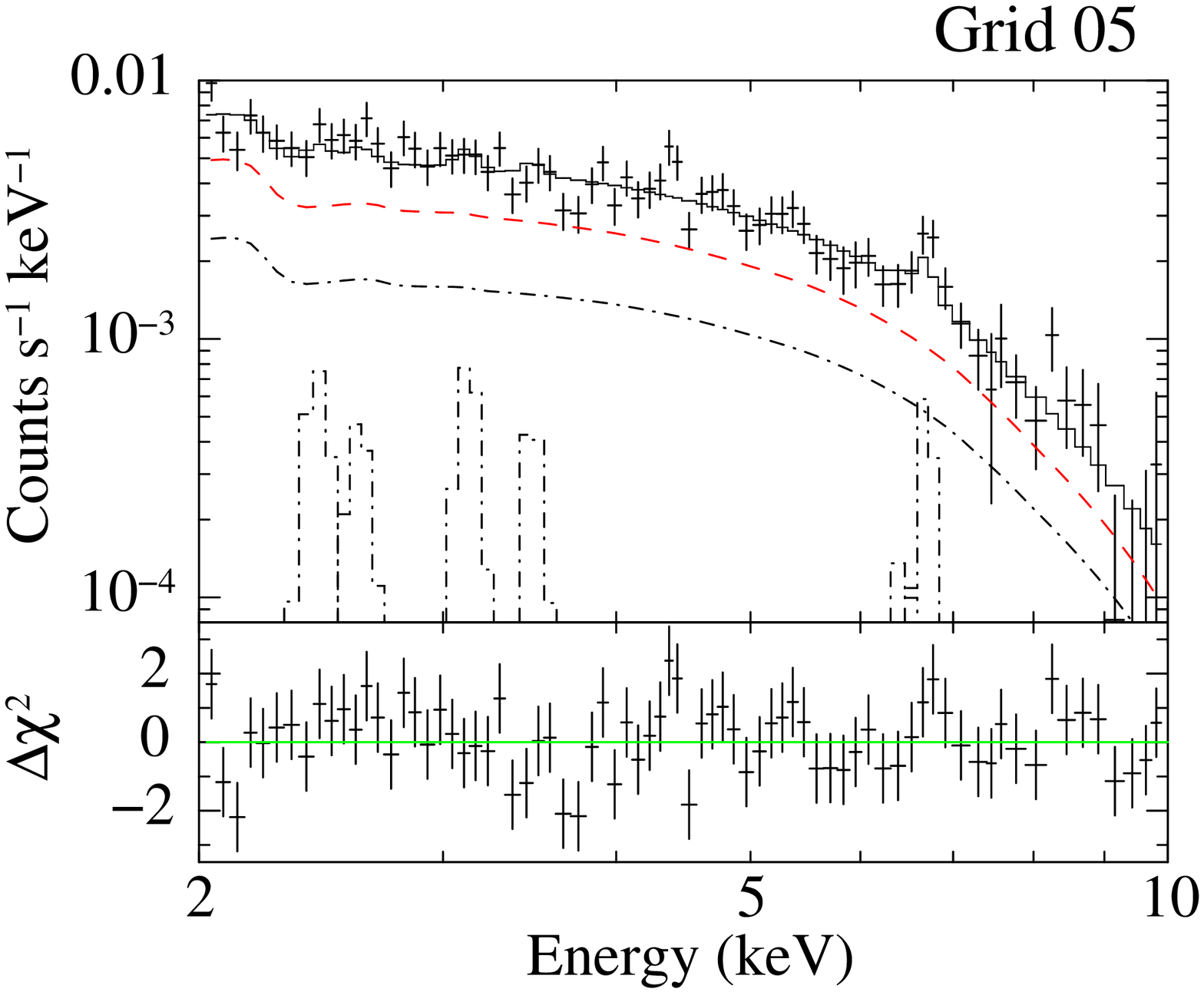}
\FigureFile(40mm,40mm){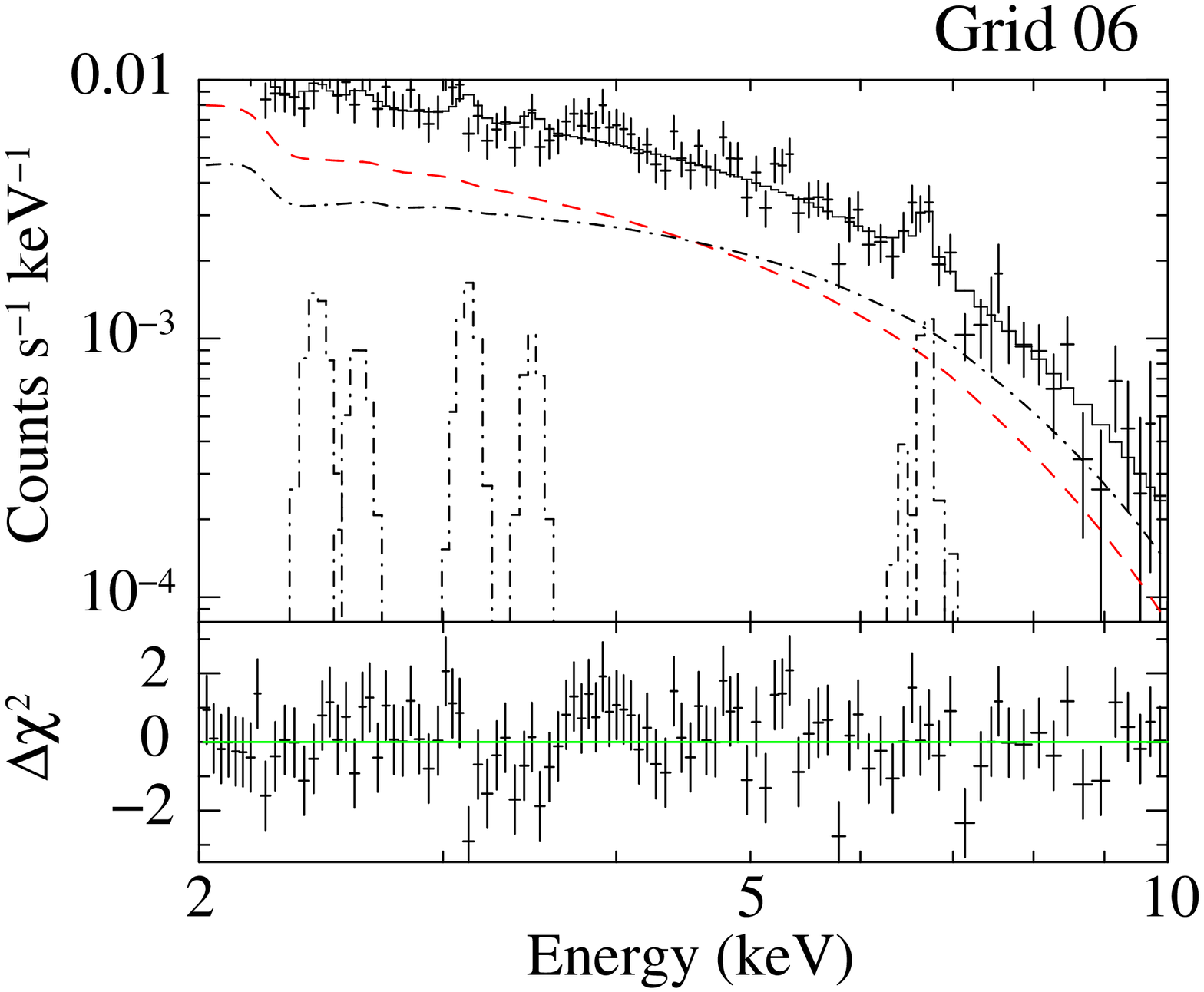}
\FigureFile(40mm,40mm){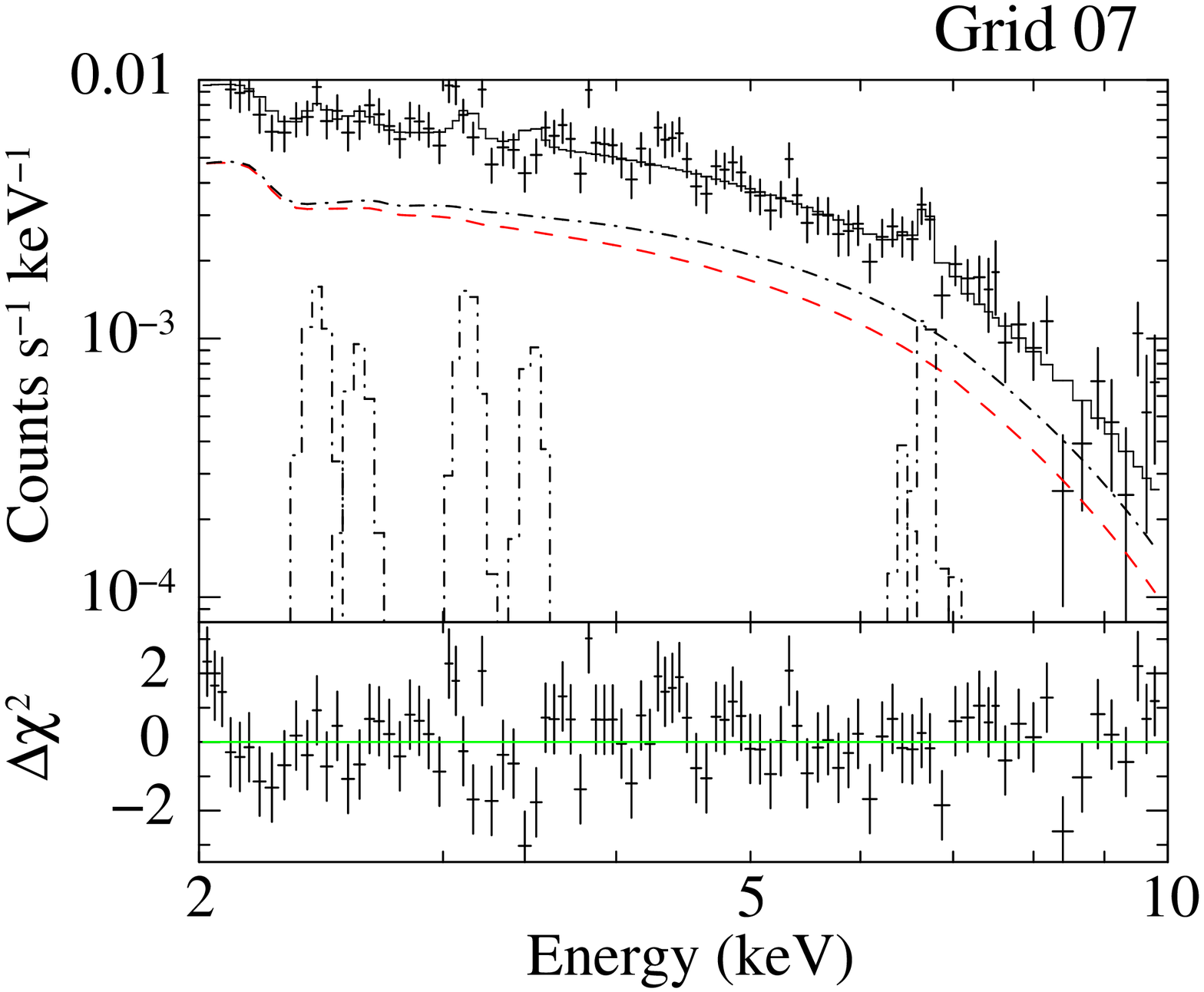}
\FigureFile(40mm,40mm){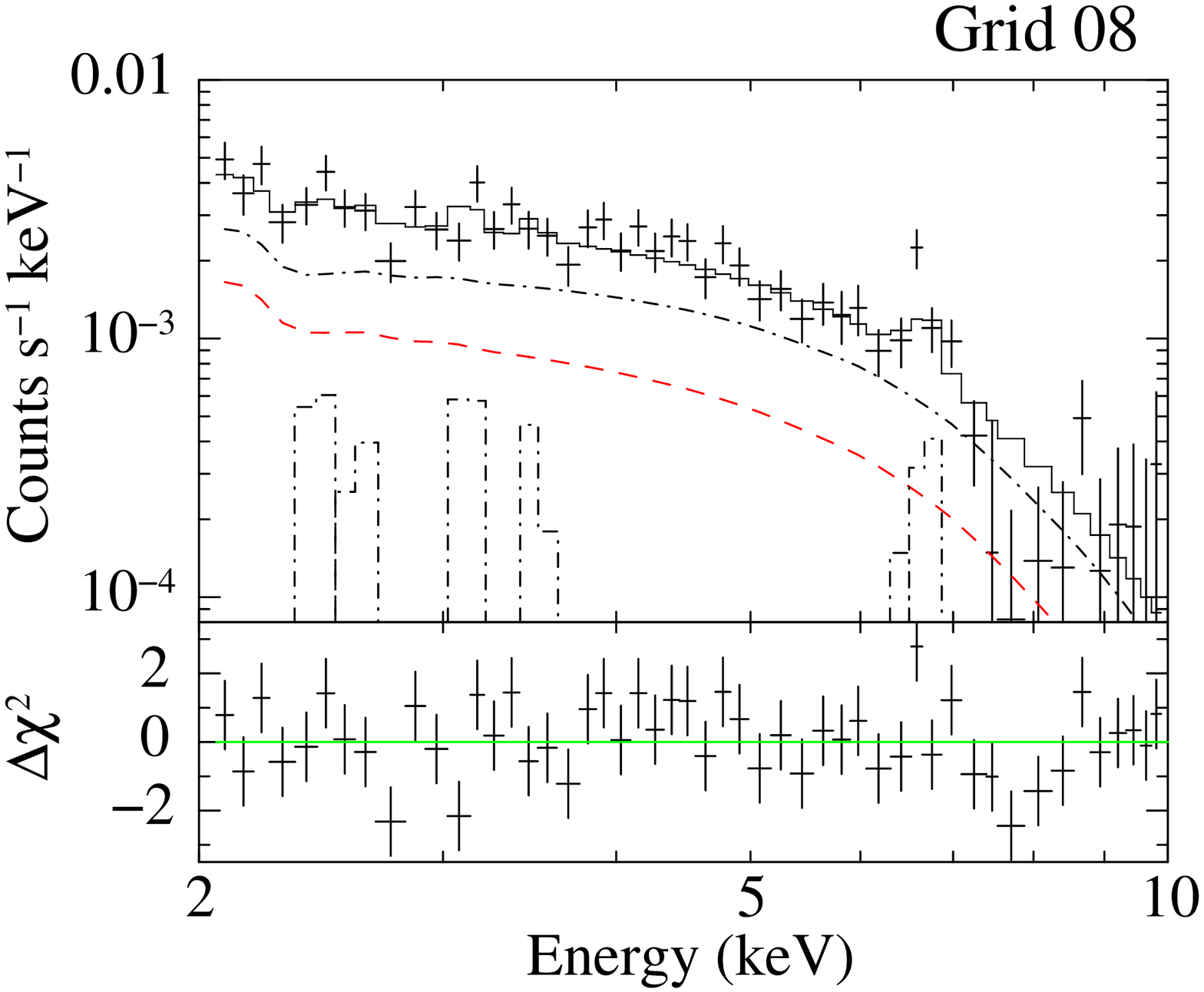}
\FigureFile(40mm,40mm){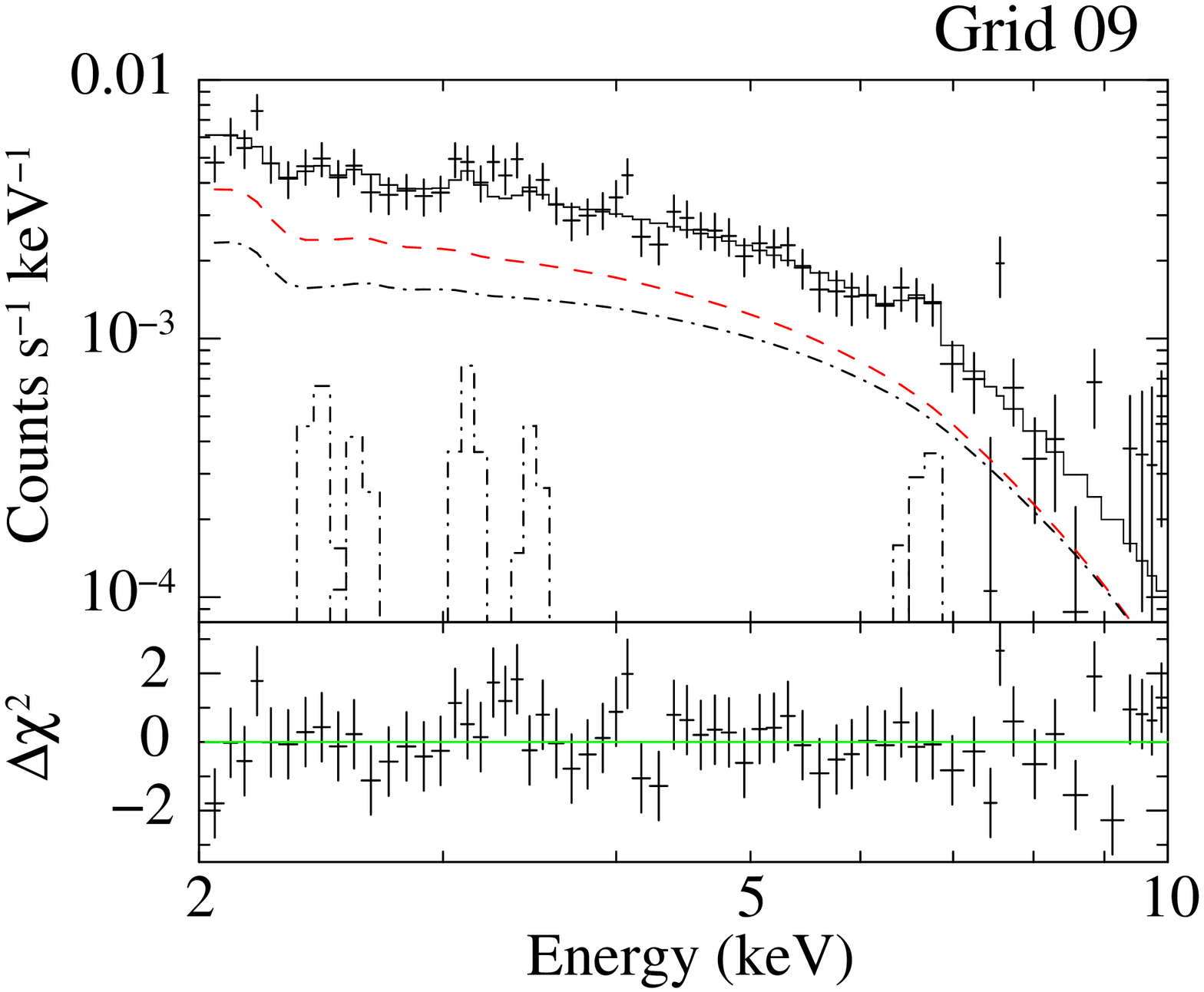}
\FigureFile(40mm,40mm){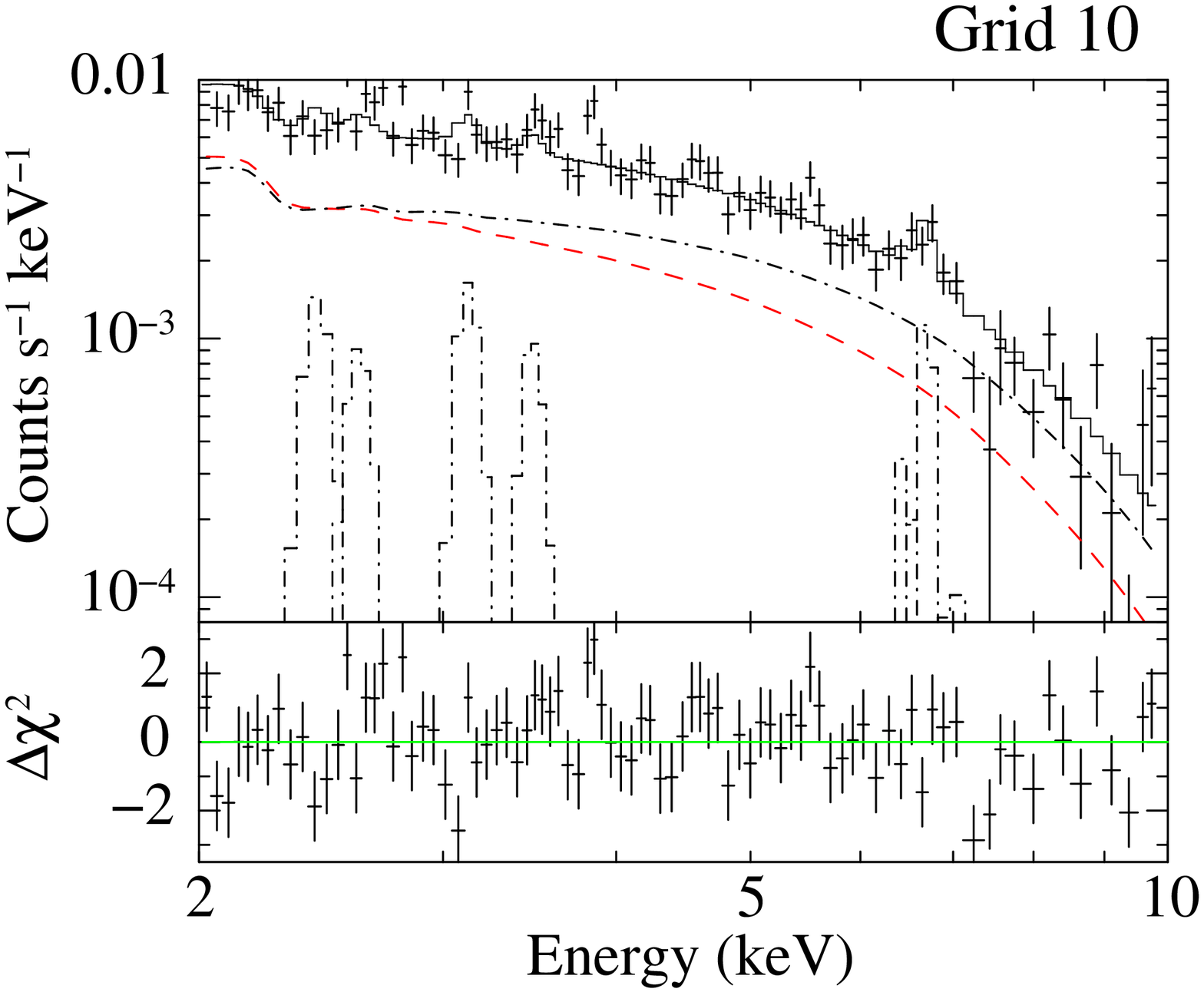}
\FigureFile(40mm,40mm){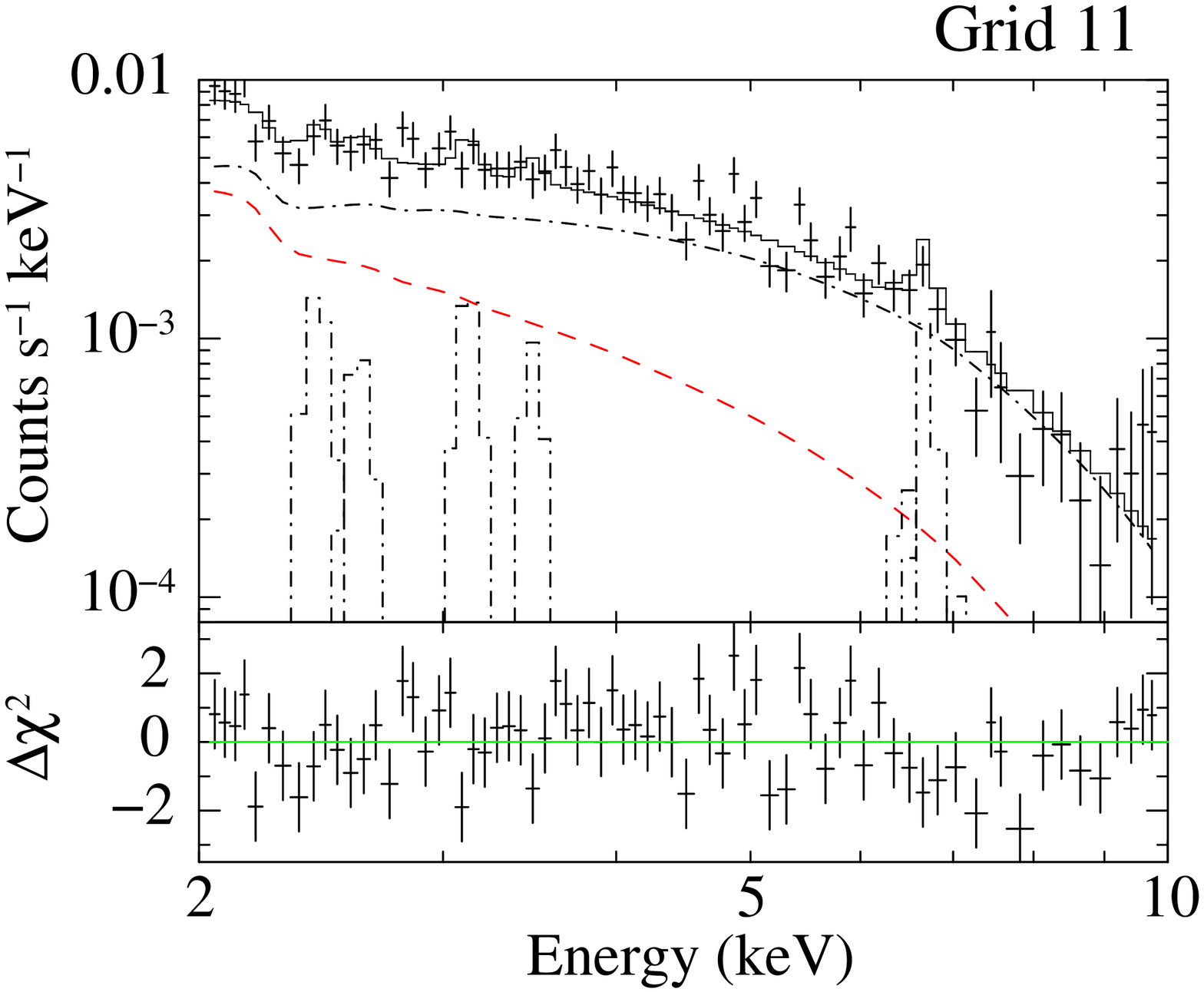}
\FigureFile(40mm,40mm){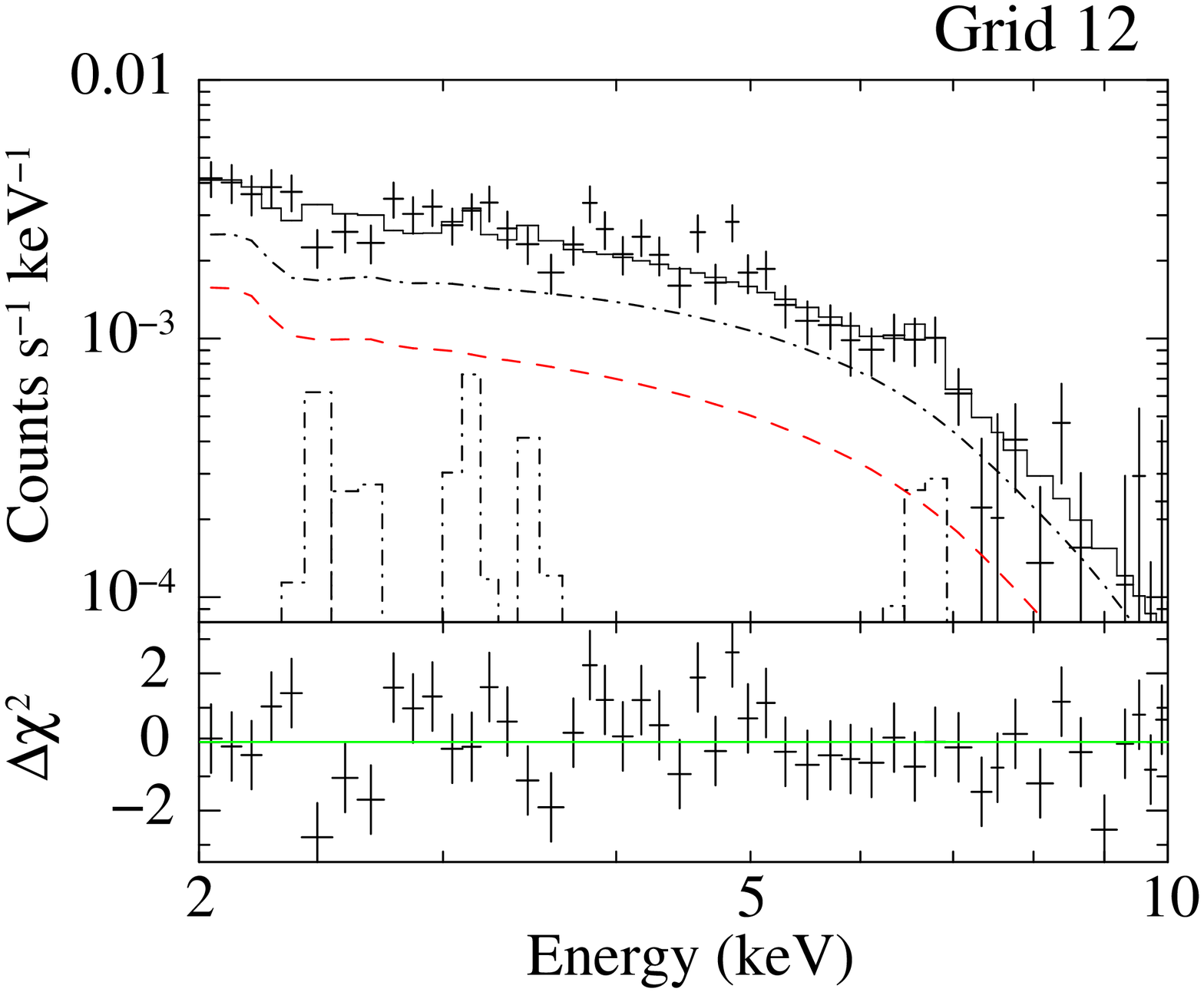}
\FigureFile(40mm,40mm){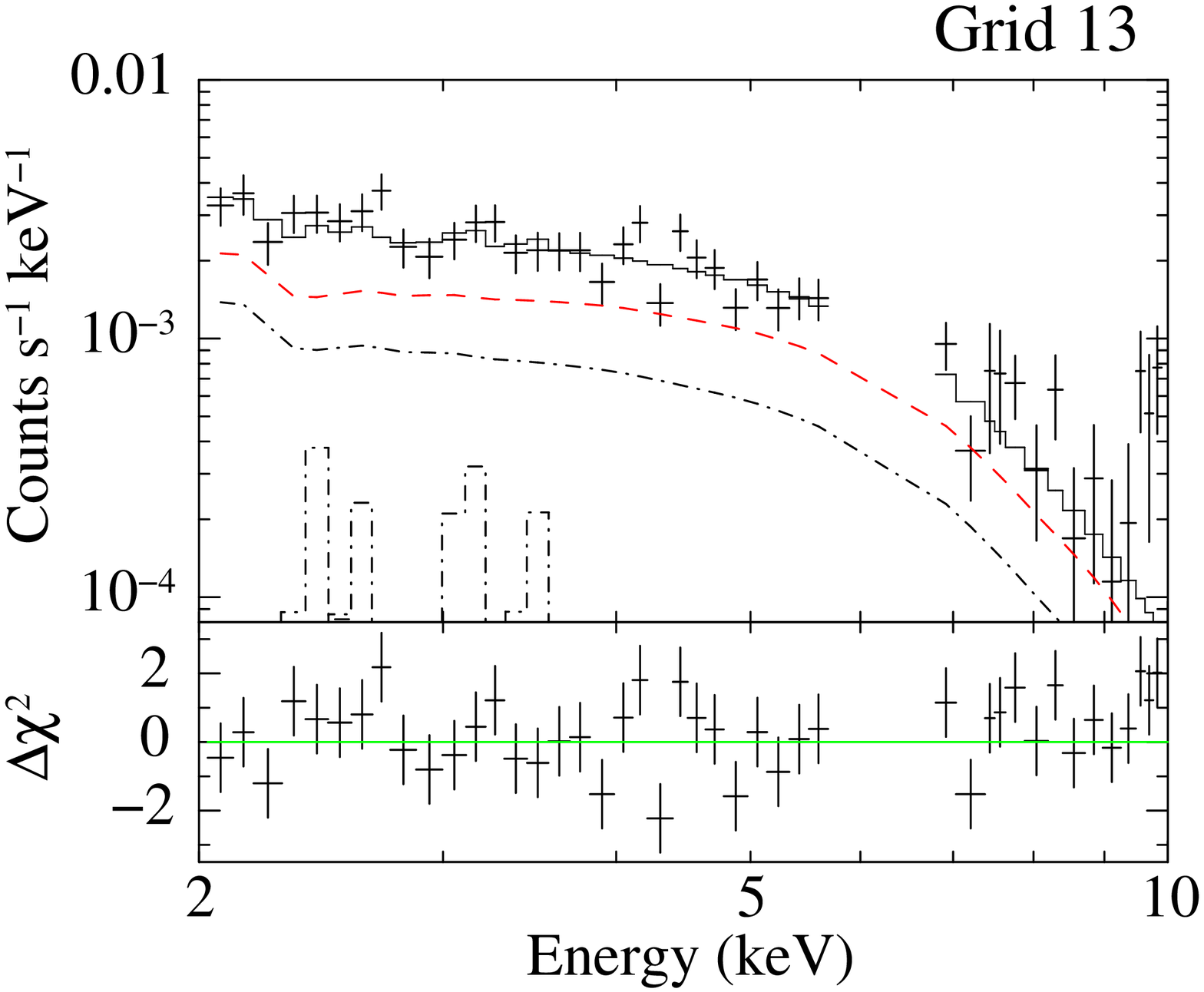}
\FigureFile(40mm,40mm){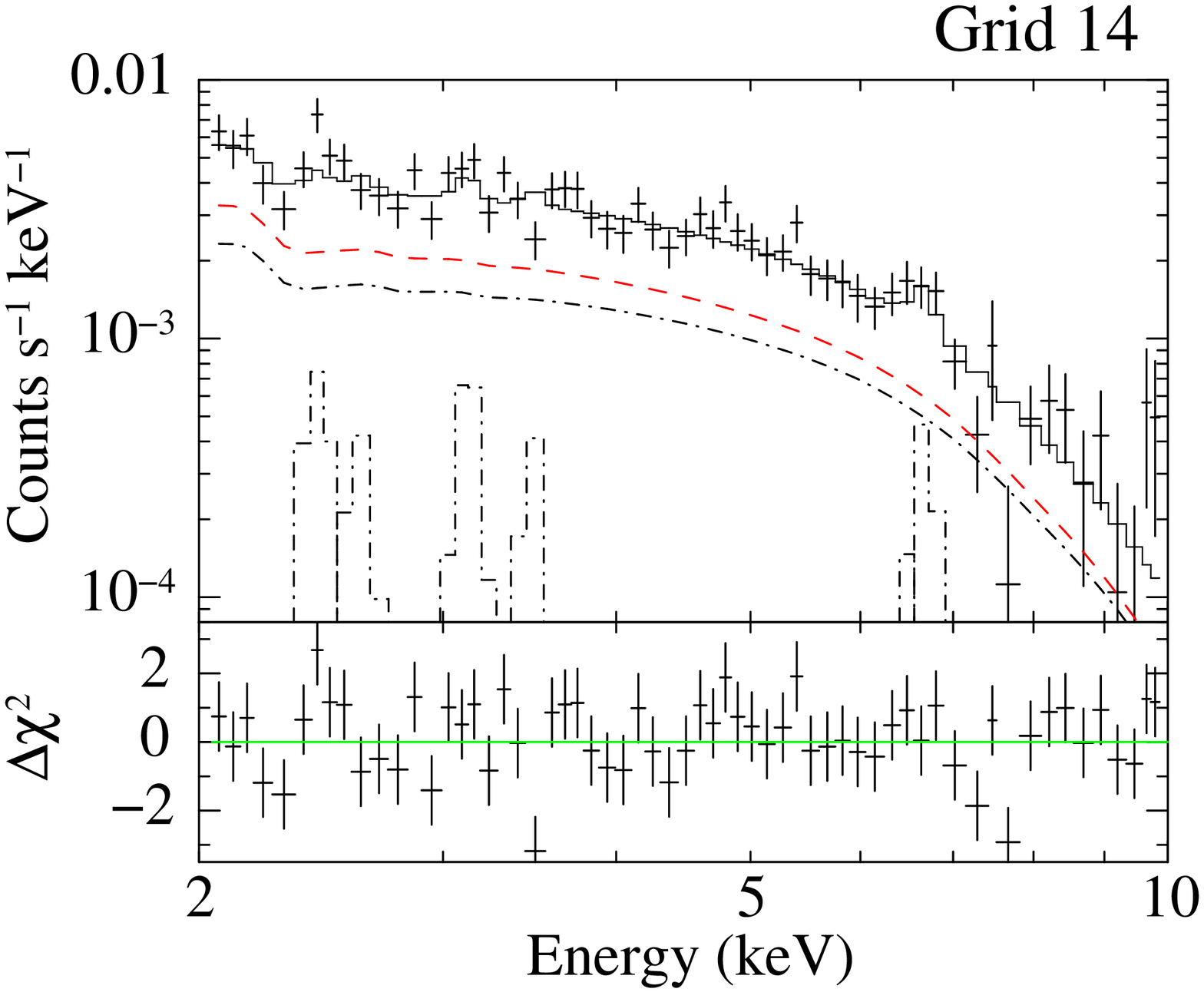}
\FigureFile(40mm,40mm){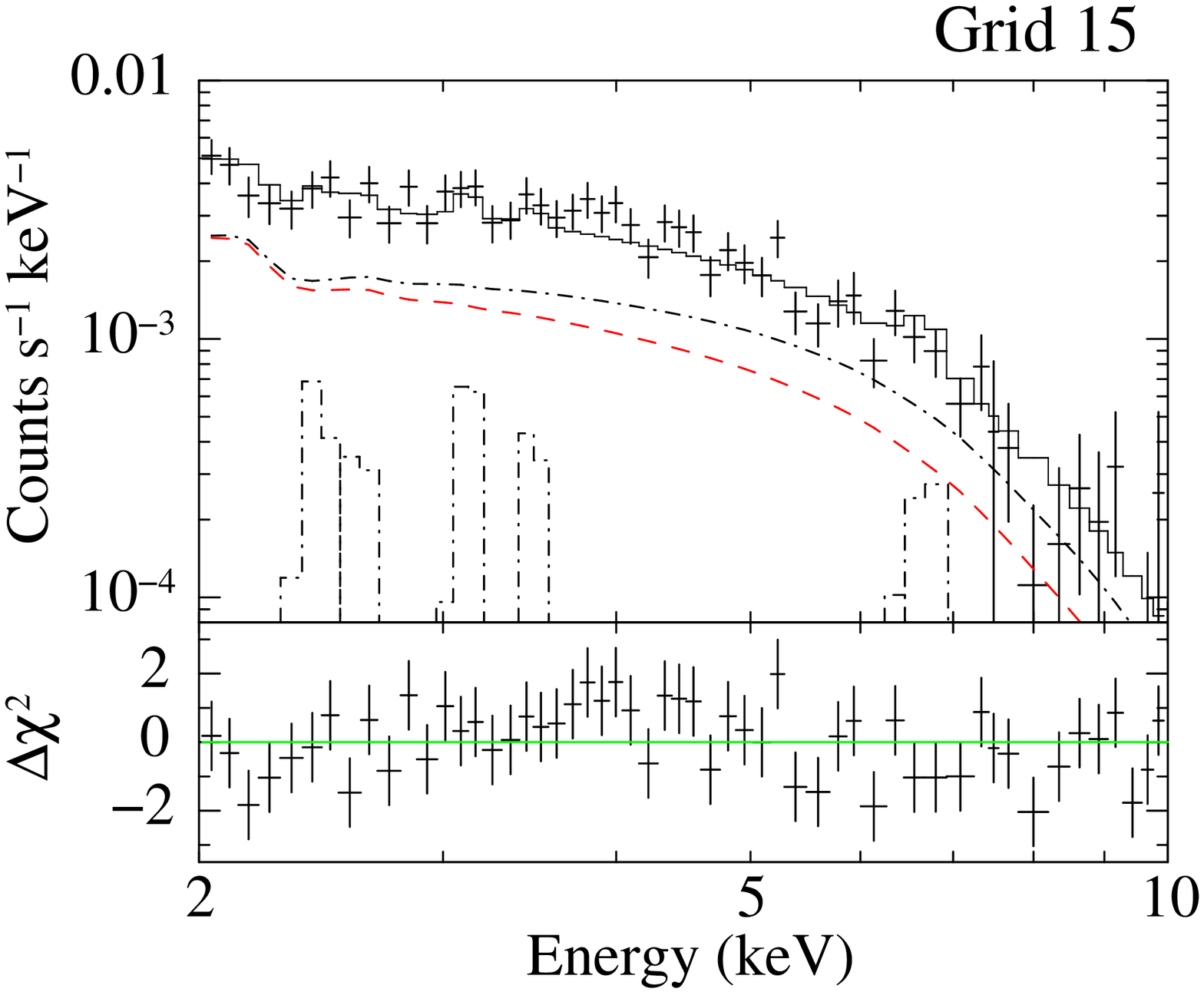}
\FigureFile(40mm,40mm){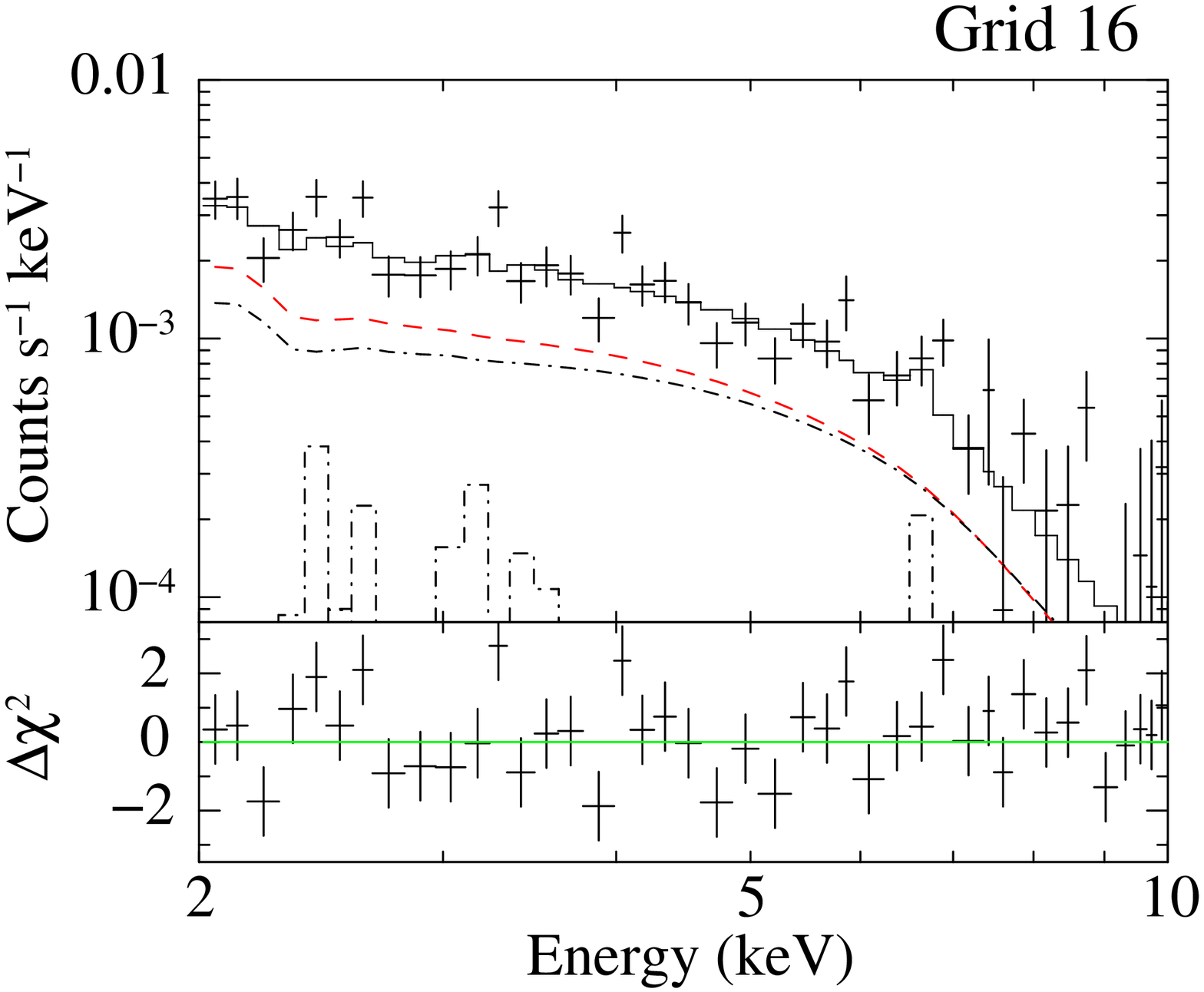}
\end{center}
\caption{
NXB-subtracted XIS spectra in the 2.0--10.0~keV band extracted from
each grid shown in
figure~\ref{fig:1809:sum013 reg 0548-2740 expcor}. Dashed line,
dash-dot line and dash-dot histograms indicate the best-fit absorbed
power-law model, a GRXE model and emission lines, respectively.
}
\label{fig:1809:multi sum013 gridall min50 fit}
\end{figure}

The model GRXE spectrum was constructed as follows:
The background spectrum was fitted in the
2.0--7.2~keV band with the model of an absorbed power-law
plus 7 narrow Gaussians as explained above. %(three for the iron lines and four for
% emission lines below 4~keV; Ebisawa et al.\ 2008).
The absorption column density was fixed to $1.0 \times 10^{22}$~cm$^{-2}$.
%since it was difficult to determine with data  above 2.0 keV\@.
The best-fit parameters,  besides
the iron line parameters in table \ref{tbl:1809:eqwidth}, are listed in table~\ref{tbl:1809:bgd spec fit}.
% We generated the GRXE spectrum model using these best-fit parameters.
% except for the overall normalization.
Normalization for each extraction region was adjusted taking account of 
differences of the vignetting effects, size of the extraction
region, and  exposure time between the north and south pointings.
The correction factors of the vignetting effect (shown in
figure~\ref{fig:1809:sum013 reg 0548-2740 expcor} right) were determined
by simulation. % ; we simulated the intensity map for the flat field with the energy spectrum of GRXE using {\tt xissim}.
We subtracted the non X-ray background (NXB) estimated by using {\tt xisnxbgen}.
% We fitted thus obtained spectra with an absorbed power-law
% plus the corrected GRXE model.
Figure~\ref{fig:1809:multi sum013 gridall min50 fit} shows the
XIS spectra (averaged for XIS 0, 1 and 3) and the best-fit models
for all the 16 regions in the 2.0--10~keV band.
Because Grid 1, 4, 13 were illuminated with the calibration sources,
the data between 5.73--6.67~keV were removed from the fit.
% As shown in these spectra, iron lines originated from GRXE were
% subtracted appropriately.
% This means that our estimation of the GRXE intensity was correct.
Fit results are summarized in table~\ref{tbl:1809:spec fit para}.
The spatial distribution of the spectral indices is shown in
figure~\ref{fig:1809:phoindex grid}.
In order to make statistics better,
we also fit the combined spectra of Grid 6 and 7
(for the region close to the emission peak)
and Grid 11, 14, and 15 
(for the region far from the emission peak).
The results are included in Table~\ref{tbl:1809:spec fit para}.
From these analyses,
we concluded that there is no systematic trend of the spatial variations
in the spectral index in this field of view.

\begin{table*}[tbp]
\begin{center}
\caption{Best-fit parameters of an absorbed power-law model with
emission lines.}
\label{tbl:1809:bgd spec fit}
\begin{tabular}{llll}
\hline
Model component & Parameter && Value \\
\hline
% \multicolumn{3}{l}{Continuum} \\
Continuum & \multicolumn{2}{l}{$N {\mathrm{H}}$
($10^{22}$~cm$^{-2}$)} & 1.0 (fixed) \\
& \multicolumn{2}{l}{$\Gamma$} & $1.40_{-0.06}^{+0.05}$ \\
% \multicolumn{3}{l}{Emission lines} \\
Emission lines & Energy (keV)
& S\emissiontype{XV} & $2.43 \pm 0.02$ \\
&& S\emissiontype{XVI} & $2.61 \pm 0.04$ \\
&& Ar\emissiontype{XVII} & $3.13 \pm 0.02$ \\
&& Ar\emissiontype{XVIII} & $3.43_{-0.07}^{+0.13}$ \\
\hline
\multicolumn{4}{@{}l@{}}{\hbox to 0pt{\parbox{180mm}{\footnotesize
Note. --- Errors represent single-parameter 90\% confidence limit.
\par\noindent
}\hss}}
\end{tabular}
\end{center}
\end{table*}

\begin{table*}[tbp]
\begin{center}
\caption{Best-fit parameters of an absorbed power-law model %plus  GRXE model (fixed) 
for each grid.}
\label{tbl:1809:spec fit para}
\begin{tabular}{cccc}
\hline
Grid id & $\Gamma$ & Flux$^*$ (10$^{-13}$~ergs~cm$^{-2}$~s$^{-1}$) & $\chi^2$/d.o.f.\\
\hline
1 & $1.52 \pm 0.21$ & $8.7 \pm 0.9$ & 52.2/46 \\
2 & $1.83 \pm 0.18$ & $7.8 \pm 0.6$ & 70.6/67	\\
3 & $1.79 \pm 0.21$ & $5.9 \pm 0.6$ & 54.1/62	\\
4 & $1.39 \pm 0.28$ & $6.1_{-0.8}^{+0.9}$ & 49.1/37 \\
5 & $1.50 \pm 0.14$ & $10.6 \pm 0.7$ & 74.0/73	\\
6$^\dagger$ & $2.07 \pm 0.14$ & $8.2 \pm 0.5$ & 105.0/102	\\
7$^\dagger$ & $1.66 \pm 0.17$ & $6.7 \pm 0.5$	& 121.7/93 \\
8 & $1.71_{-0.38}^{+0.40}$ & $3.0 \pm 0.5$ & 59.2/48	\\
9 & $1.71 \pm 0.18$ & $7.1 \pm 0.6$ & 58.1/61	\\
10 & $1.94 \pm 0.18$ & $5.9 \pm 0.5$ & 122.9/86	\\
11 & $2.74_{-0.35}^{+0.39}$ & $2.4 \pm 0.4$ & 87.6/70	\\
12 & $1.72_{-0.37}^{+0.38}$ & $2.9 \pm 0.5$ & 65.2/47	\\
13 & $1.19 \pm 0.26$ & $8.3 \pm 1.0$ & 51.4/41	\\
14 & $1.54 \pm 0.20$ & $7.2_{-0.6}^{+0.7}$ & 74.3/59 \\
15 & $1.78 \pm 0.25$ & $4.4_{-0.5}^{+0.6}$ & 55.5/52 \\
16 & $1.66 \pm 0.32$ & $4.9 \pm 0.7$ & 60.8/40 \\
6$^\dagger$+7$^\dagger$ & $1.91 \pm 0.11$ & $14.7 \pm 0.9$ & 235.9/196 \\
11+14+15 & $1.84\pm 0.14$ & $13.7_{-1.4}^{+1.5}$ & 243.2/183 \\
\hline
\multicolumn{3}{@{}l@{}}{\hbox to 0pt{\parbox{120mm}{\footnotesize
Note. --- Errors represent single-parameter 90\% confidence limit.
\par\noindent
$^*$ Unabsorbed flux in the 2.0--10.0~keV band.
\par\noindent
$^\dagger$ The emission is dominated by the point source.
}\hss}}
\end{tabular}
\end{center}
\end{table*}

\section{Discussion}
\label{sec:discussion}

Using Suzaku, we have clearly shown  presence of the large scale diffuse emission 
around HESS~J1809--193, 
which was already suggested by ASCA \citep{2003ApJ...589..253B}.
The extension of the diffuse emission is  at least $\sim\!21'$
($3\sigma$ of the gaussian approximation).
The diffuse spectrum has the photon index of $\Gamma \sim$1.7,
which is much harder than those of SNRs with synchrotron X-ray emitting shells
($\Gamma \sim$2--3),
such as SN~1006, RX~J1713--3946, and others
\citep[for example]{bamba2008,takahashi2008}.
Such a hard spectrum, on the other hand,
reminds us the PWN origin \citep{kargaltsev2008}.
This hypothesis is, in fact, strengthened by the very existence of the pulsar PSR~J1809$-$1917.

The TeV emission also may have the   PWN origin,
because it positionally coincides with the region including 
the pulsar and the diffuse X-ray emission,
although the center of the TeV emission is offset by
$\sim\!6'$ (a projected distance of 6~pc at 3.5~kpc) from the pulsar.
If both the TeV and X-ray emissions are from the same PWN,
their origin is accelerated electrons via inverse Compton (TeV)
and synchrotron mechanisms (X-ray).
Typical energies of responsible electrons are $\sim$20~TeV for the TeV emission
and $\sim$80~TeV for X-rays, respectively.
The former has much longer synchrotron life timescales than latter
\citep[for example]{2009ApJ...694...12M}.
If we assume that the electrons which emit TeV gamma-rays
have the same age as the pulsar ($\tau c = $51~kyr)
and those for the X-rays are very fresh, the positional offset 
may be explained by the proper motion of the pulsar.
To explain the offset of $\sim\!6'$, the transverse velocity of the pulsar 
needs to be $\sim$120~km~s$^{-1}$,
which can be explained with the average transverse velocity of
radio pulsars
\citep[$\sim 300$~km~s$^{-1}$]{1994Natur.369..127L}.
%However, the pulsar tail detected by {\it Chandra}
%\citep{2007ApJ...670..655K} 
%shifts $\sim 60 - 70$~degrees from the offset direction,
%which makes this scenario unlikely.
{\bf
On the other hand,
the tails seen in the {\it Chandra} image (if it is indeed a tail not a jet)
suggests that the pulsar is moving southward and its velocity vector is
at the angle of $\sim$190--200~deg. East of North
\citep{2007ApJ...670..655K}.
Now if we assume that the pulsar was born at the center of the TeV
source and moved to its current position,
its velocity vector would have
to be at the angle of 10--20~deg. West of North.
Therefore, the angle between these
two velocity vectors will be $\sim$140--160~deg.,
which makes this scenario unlikely.
}
Another possible scenario to make such an offset is collision between the
reverse shock and the PWN
\citep{2008ApJ...675..683P}
like the case for Vela~X \citep{2001ApJ...563..806B,2003ApJ...588..441G}.

In the young PWNe, it is suggested that
the X-ray spectra become softer with the distance from the pulsars
due to synchrotron energy loss of the accelerated electrons
\citep[for example]{mori2004}.
% thus we could see it in our target.
However, we could not find any hint of such softening
(Figure~\ref{fig:1809:phoindex grid}).
This means that electrons do not lose significant energies 
via the synchrotron emission.
The X-ray size of HESS~J1809--193 is $\sim\!21'$ ($=3\sigma$),
or 21~pc with the assumed distance of 3.5~kpc.
The synchrotron lifetime of an accelerated electrons ($\tau {syn}$) is
$\tau {syn} = 6.8{\rm kyr}(B/3~\mu{\rm G})^{-3/2}(E {syn}/2~{\rm keV})^{-1/2}$,
where $B$ and $E {syn}$ are
magnetic field and mean energy of synchrotron emission
from accelerated electrons, respectively. 
In order to explain the size of the X-ray diffuse emission,
the transport velocity of accelerated electrons should be higher than
21~pc/$\tau {syn} \sim 3.0\times 10^3 \;{\rm km \; s^{-1}} \; (B/3~\mu{\rm G})^{3/2}(E {syn}/2~{\rm keV})^{1/2}$,
which seems to be very fast for an old PWN, 
and even comparable to the forward shock velocities of young SNRs.
It is known that the diffusion coefficient in young PWNe and SNRs
is too small to diffuse out to such a large scale,
when the magnetic field is turbulent 
\citep[for example]{shibata2003,2003ApJ...589..827B,2005ApJ...621..793B}.
Therefore, the current observation suggests that 
the turbulence of magnetic fields in old PWN systems are smaller than
those in young systems.
If turbulence of the magnetic fields in old PWN systems is smaller than
that in young systems, such a fast diffusion may  be explained,
although we have no such a model for old PWNe.
In supernova remnants, we have some indication that
the turbulence becomes smaller when the SNR becomes older
\citep{2005ApJ...621..793B},
similar mechanism might work for PWNe.

We have a similar case, HESS~J1825--137,
which is another unID TeV source with   possible PWN origin.
\citet{2009PASJ...61S..189T} found that the X-ray photon index is
nearly constant over the emission region, which may suggest 
HESS~J1825--137 also might have small turbulence of the magnetic field.
Therefore, we suppose  fast transportations of accelerated  electrons may be
rather common phenomena in old PWNe. 
Further similar samples of unID TeV sources
may confirm this hypothesis.

\section{Summary}
\label{sec:1809:summary}

\begin{itemize}
\item
We observed the TeV PWN candidate HESS J1809--193 with Suzaku, and
confirmed an  extended emission  around the pulsar PSR~J1809--1917.
Size of the X-ray emission is at least $\sim 21'$,
or 21~pc at 3.5~kpc.
\item
The extended emission has very hard nonthermal spectrum
with the photon index of $\sim$1.7.
No systematic spatial variation of the photon index is found,
which implies that 
accelerated electrons do not lose their energy
when they run from the pulsar to the edge of the emission region.
We have to consider very fast diffusion % or another mechanism
in an old PWN to reproduce such phenomena. 
If turbulence of the magnetic fields in old PWN systems is smaller than
that in young systems, such a fast diffusion may  be explained.
\end{itemize}

\section*{Acknowledgements}

We would like to thank the anonymous referee for useful comments
and suggestions.
We acknowledge all the Suzaku team members for their gracious supports.
The authors also thank K.~Mori and R.~Yamazaki
for their fruitful comments.
A.~Bamba is supported by JSPS Research Fellowship for Young Scientists
(19-1804).

\end{document}